\documentclass[sn-nature]{sn-jnl}

\usepackage{graphicx}%
\usepackage{multirow}%
\usepackage{amsmath,amssymb,amsfonts}%
\usepackage{amsthm}%
\usepackage{mathrsfs}%
\usepackage[title]{appendix}%
\usepackage[table,xcdraw]{xcolor}
\usepackage{textcomp}%
\usepackage{manyfoot}%
\usepackage{booktabs}%
\usepackage{algorithm}%
\usepackage{algorithmicx}%
\usepackage{algpseudocode}%
\usepackage{listings}%
\usepackage{comment}
\usepackage{multicol}
\usepackage{multirow}

\usepackage{setspace}

\begin{document}

\title[Defaults: a double-edged sword in governing common resources]{Defaults: a double-edged sword in governing common resources}



\author*[1,2,3]{\fnm{Eladio} \sur{Montero-Porras}}\email{eladio.montero@vub.be}

\author[4]{\fnm{Rémi} \sur{Suchon}}\email{remi.suchon@univ-catholille.fr}

\author[1,2,3,5]{\fnm{Tom} \sur{Lenaerts}}\email{tom.lenaerts@ulb.be}

\author*[1,2,3]{\fnm{Elias} \sur{Fernández Domingos}}\email{elias.fernandez.domingos@ulb.be}

\affil[1]{\orgdiv{Artificial Intelligence Laboratory}, \orgname{Vrije Universiteit Brussel}, \orgaddress{\street{Pleinlaan 9}, \city{Elsene}, \postcode{1050}, \state{Brussels}, \country{Belgium}}}

\affil[2]{\orgdiv{Machine Learning Group}, \orgname{Université Libre de Bruxelles}, \orgaddress{\street{Boulevard du Triomphe CP212}, \city{Elsene}, \postcode{1050}, \state{Brussels}, \country{Belgium}}}

\affil[3]{\orgdiv{FARI Institute}, \orgname{Universite Libre de Bruxelles-Vrije Universiteit Brussel}, \city{Brussels}, \postcode{1000}, \country{Belgium}}

\affil[4]{\orgdiv{Anthropo-Lab - ETHICS}, \orgname{Université Catholique de Lille}, \orgaddress{\street{Maison des chercheurs, 14 Bd Vauban}, \city{Lille}, \postcode{59000}, \country{France}}}

\affil[5]{\orgdiv{Center for Human-Compatible AI}, \orgname{University of California, Berkeley}, \orgaddress{\street{2121 Berkeley Way}, \city{Berkeley}, \postcode{94720}, \state{CA}, \country{USA}}}

\doublespacing

\abstract{Extracting from shared resources requires making choices to balance personal profit and sustainability. We present the results of a behavioural experiment wherein we manipulate the default extraction from a finite resource. Participants were exposed to two treatments --- pro-social or self-serving extraction defaults --- and a control without defaults. We examined the persistence of these nudges by removing the default after five rounds. Results reveal that a self-serving default increased the average extraction while present, whereas a pro-social default only decreased extraction for the first two rounds. Notably, the influence of defaults depended on individual inclinations, with cooperative individuals extracting more under a self-serving default, and selfish individuals less under a pro-social default. After the removal of the default, we observed no significant differences with the control treatment. Our research highlights the potential of defaults as cost-effective tools for promoting sustainability, while also advocating for a careful use to avoid adverse effects.}




\maketitle

\section{Introduction}\label{sec1}

Modern life relies heavily on finite resources like water, electricity, internet bandwidth or public roads. The over-consumption of these resources may produce negative externalities on society and the environment. These pose significant challenges and are difficult to overcome, as individual rationality may easily lead to over-consumption, resulting thus in a ``tragedy of the commons'' \cite{hardin_tragedy_1968}. Resolving such social dilemmas require the design of institutions and mechanisms to enforce rules. A central authority, e.g. the state, can impose sanctions and rewards (such as taxes, or extraction rights) to tame over-consumption. In opposition to such a top-down approach, Elinor Ostrom's seminal work \cite{ostrom_rules_1994} demonstrates the effectiveness of decentralised, bottom-up arrangements in preserving exhaustible resources through self-governance.  

The consumption of finite resources often occurs through provider companies (e.g., water and energy suppliers) or governmental institutions (e.g., natural resource management agencies, environmental protection departments), to enumerate some examples. These mediating entities can play a decisive role in curbing over-consumption, by encouraging participants towards more desirable consumption behaviours. For example, energy providers offer diverse consumption plans tailored to their capacities and users' schedules and preferences, including selecting energy from environmentally friendly or cost-effective sources. The same can be found for other resources, such as water or internet access \cite{abrahamse_review_2005, momsen_intention_2014}. 

As the number of provided options grows, intelligent automated systems become useful to assist in making choices. In the context of energy consumption, for example, automated systems can schedule when household appliances become active or when hybrid cars should charge during off-peak hours, linking these choices to when sustainable energy sources are providing more energy than non-sustainable ones \cite{setlhaolo_optimal_2014}. An important issue among these decision-support systems is that their design is determined by the mediating entities mentioned earlier. They can predefine a set of default consumption plans for users, determined for instance by assumed users' preferences or needs, their own interest or broader prosocial goals, without reducing freedom of choice, while also being easy to implement \cite{jachimowicz_when_2019}. 

Users may respond differently to the available default setting,  influenced by their own inclinations and priorities \cite{fischbacher_social_2010}. In this line, Behlen et al. found that defaults, while being effective, require a targeted approach to reach individuals whose interests align with the policy-maker \cite{behlen_defaults_2023}. For instance, an electricity provider can offer renewable energy for consumers by default, promoting a course of action supported by policy or a prevailing social norm. Yet, some consumers may prefer other sources of energy and adapt their consumption habits to accommodate their individual preferences  \cite{taube_how_2019, liebe_large_2021}. For instance, a person who is particularly insensitive to the collective issue of energy over-consumption may be more likely to override a default directed towards renewable energy sources. So far, no clear answers have been provided experimentally on this association between personal preferences, default settings and consumption of a finite resource. In this work, we aim to address this gap in the literature by examining this question. 

\begin{figure}
    \centering
    \includegraphics[width=1\linewidth]{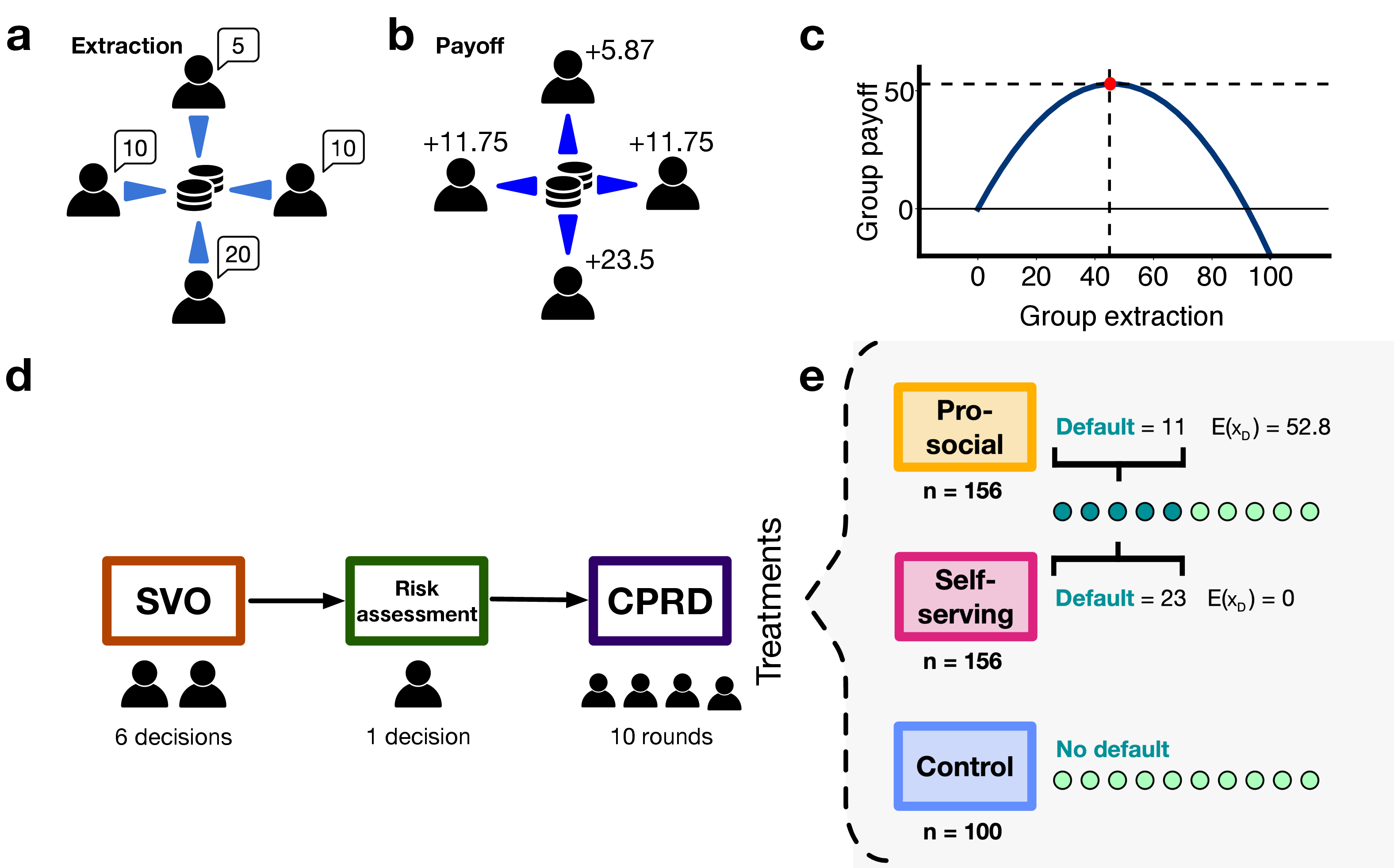}
    \caption{\textbf{Visual representation of the experiment designed for this work}. We use the Common Pool Resource dilemma (CPRD) \cite{walker_rent_1990} to understand collective resource management. In this game, individuals extract resources (as tokens) and receive payoffs (in experimental units, or ECUs) proportionate to their extraction levels. Panel \textbf{A} shows four individuals who request 45 tokens in total. Panel \textbf{B} transforms these requests into a collective payoff of 52.87 ECUs, determined by a payoff function curve shown in panel \textbf{C}. (more details in the Methods section). \textbf{D: } Experiment flow. the experiment consists of three tasks: In task 1 (brown box), participants complete an incentivised Social Value Orientation task (SVO) \cite{murphy_social_2013}, allowing us to identify participants' resource allocation preferences. Task 2 (green box), assessed participants' risk aversion (also incentivised) \cite{dave_eliciting_2010, eckel_sex_2002}. Finally, in task 3, participants engage in the CPRD as part of one of three treatments, which cover 2 different types of defaults and one treatment without a default for comparison (panel \textbf{E}): Pro-social (default = 11 tokens, optimal group extraction yielding 52.8 ECUs), Self-serving (default = 23, excessive extraction yielding no returns), and a Control with no preset extraction. The green dots represent the rounds, where the darker green represent the first five rounds with default values, and the lighter green the rounds with no default. More details on these treatments are provided in the Methods section.}
    \label{fig:experiment_flow}
\end{figure}

First, we want to assess whether a simple manipulation, like setting a default extraction value, can curb collective (over-)consumption in a Common Pool Resource dilemma (CPRD) \cite{walker_rent_1990, ostrom_rules_1994}. The CPRD is particularly well-suited to study consumption of a finite resource: CPRD experiments have been used to study the exploitation of water basins, lakes, irrigation of a community, fisheries or timber, to name a few \cite{alpizar_effect_2010, bernedo_del_carpio_community-based_2021, ostrom_beyond_2010, schill_sustaining_2023}. The problem lies in the aggregate behaviour of participants: without a system of governance, participants excessively appropriate from the common resource, which may give rise to a "tragedy of the commons" \cite{falk_appropriating_2001}. See Figure \ref{fig:experiment_flow}A for a visual representation of the CPRD.

A meta-analysis by Mertens et al. demonstrated that choice architecture strategies, particularly those on decision structure like default settings, often surpass others focusing on decision information or assistance \cite{mertens_effectiveness_2022}, revealing the potential efficacy of such interventions. Fosgaard et al. observed that in public goods games, presenting a conditional cooperation strategy as the default option effectively nudged participants towards more cooperative behaviour \cite{fosgaard_nudge_2015}. Similarly, Bynum et al. studied how default non-participation in collective risk dilemmas (contributing zero by default) led to participants more frequently failing to meet the required threshold in this game\cite{bynum_passive_2016}. Moreover, Ferguson et al. found in their dynamic organ donor game that default opt-out decisions for non-cooperators significantly impact group cooperation levels, more so than opt-in decisions for cooperators \cite{ferguson_when_2020}. Furthermore, default settings significantly impact environmental choices. Liebe et al. showed that when sustainable, or 'green', energy options are set as the default, their adoption rates increase, resulting in reduced energy consumption \cite{liebe_large_2021}. Conversely, when 'grey' energy options, which are less eco-friendly, are defaulted, participants adopted these less sustainable sources more often \cite{pichert_green_2008}. This study builds on prior research by evaluating defaults in the CPRD management, contrasting a socially beneficial default with a socially detrimental one and their impacts on overextraction.

Second, we aim to understand how default settings impact long-term consumption habits, and what happens when the default is lifted. This includes assessing the potential for both positive and negative spillovers. Positive spillover refers to scenarios where a pro-social behaviour leads to more of such behaviour, enhancing thus sustainability efforts. Conversely, negative spillover, or backfire effects, occur when an initial pro-social action is followed by opposing behaviours (see the work of Truelove et al. discussing these spillover effects in detail \cite{truelove_positive_2014}). 

In this regard, the experimental findings are mixed. Cappelletti et al. found a decline in choice of the default contribution following the removal of defaults in a public goods game, highlighting the challenges in sustaining behaviours when the default is removed \cite{cappelletti_are_2014}. Fosgaard et al. found that after participants saw a free-rider strategy by default, they were significantly more defective in subsequent public goods game without defaults \cite{fosgaard_nudge_2015}.
However,  contrasting these findings, Ghesla et al. found that encouraging pro-social decisions through choice defaults, with or without significant opt-out costs, does not affect unrelated subsequent pro-social behaviour \cite{ghesla_nudge_2019} and they might be effective for subsequent tasks \cite{van_rookhuijzen_effects_2021}. Albeit these soft interventions are very popular for policymaking, their long-term efficacy has been questioned \cite{lemken_public_2023}. Our work expands on this current understanding by testing the presence and nature of spillover effects when default values are lifted. 

It remains unclear how default settings, known to limit individualistic behaviour, might also have negative social impacts. Prior studies show that consumers often select products with costly, preselected add-ons by default, rather than customising basic products with their preferred options \cite{park_choosing_2000}. This pattern, evident across various shopping platforms and devices, leads to unintended purchases or non-consensual sharing of personal data due to default settings \cite{bosch_tales_2016}.
Our study examines the dual nature of default effects by introducing both pro-social and self-serving defaults, to assess their influence on group decisions towards either beneficial or harmful outcomes.

Therefore, the third and final goal is to explore how the impact of defaults on \mbox{(over-)} consumption is mediated by (the heterogeneity of) individual preferences. We introduce two different default values, which might encourage pro-social individuals towards increased extraction and individualistic ones towards reduction. Understanding the heterogeneous effect of nudges could help tailor nudge interventions to individual characteristics and make them (more) effective \cite{sunstein_nudges_2016, behlen_defaults_2023}. However, mismatches between these two can lead to backfiring, surpassing the impact of transparent information or modes of thinking \cite{de_ridder_nudgeability_2022}. Guido et al.'s experiment showed that rule-followers responded more to nudges with persuasive, socially conscious messages than rule-breakers, but no average effect was noted when social preferences were ignored  \cite{guido_nudging_2023}. Additionally, Ghesla et al. found in the context of energy consumption that politically left-leaning individuals or those prioritising environmental concerns are likelier to opt out of default gray energy contracts \cite{ghesla_behavioral_2017}.

To achieve our 3 goals, we introduce defaults in the CPRD, and we also collect data on participants' social preferences. Importantly, behaviour in such experimental social dilemmas predicts real-life behaviour, which ensures the external validity of our approach and results \cite{fehr_field_2011, rustagi_conditional_2010}. 

In the CPRD experiment, as represented in Figure \ref{fig:experiment_flow}, participants in groups of four have to coordinate their consumption of a finite resource, which offers benefits to consumers based on their proportional usage. They repeat this task for 10 rounds. Two default scenarios are presented: i) the \emph{Self-serving} default, while enticing since it leads to higher payoff if the rest extract reasonably, selecting this value results in zero benefits if it is chosen by everyone in the group; and ii) the \emph{Pro-social} default, which, if collectively adopted, yields the social optimum, maximizing the resource's potential benefits (see Methods for details). Although the default value in the CPRD suggest a course of action,  participants can override it and choose another consumption level (opt-out). This preserves the participants' autonomy over the shared resource. To assess the ability of a default to build lasting consumption habits, our second goal, we provide the default only during the first five rounds of the experiment, and remove it in the last five. 

Before the actual CPRD experiment (see Figure \ref{fig:experiment_flow}), we use a Social Value Orientation (SVO) test \cite{bogaert_social_2008} to classify participants into four social-preference types: altruistic, pro-social, individualistic and competitive individuals. Pro-social individuals, identified by their SVO, have been shown to exhibit heightened concern for environmental causes and collective welfare \cite{lange_social_1998,fleis_social_2019,cremer_why_2001}. Research has demonstrated that individuals classified as individualistic and competitive tend to extract significantly more than those classified as altruistic or prosocial in resource dilemmas \cite{roch_effects_1997}. Moreover, certain decisions can pose greater challenges for specific individuals, such as the relatively slower evidence collection among equality seekers compared to categorical decision-makers \cite{montero-porras_fast_2022}.
Lastly, we also measure the level of risk-aversion of participants and correlate it with default adoption, as the default option is often perceived as the safe and risk-free option \cite{giuliani_joint_2023}. Moreover, we want to test the relationship between risk perception and consumption, since risk-averse individuals have shown to consume less in CPRD experiments \cite{buckley_demand_2018}.


\section{Results}\label{sec:results}

\subsection{While defaults influence participants, the effect is asymmetrical \label{sec:results_1}}

\begin{figure}[!ht]%
\centering
\includegraphics[width=\linewidth]{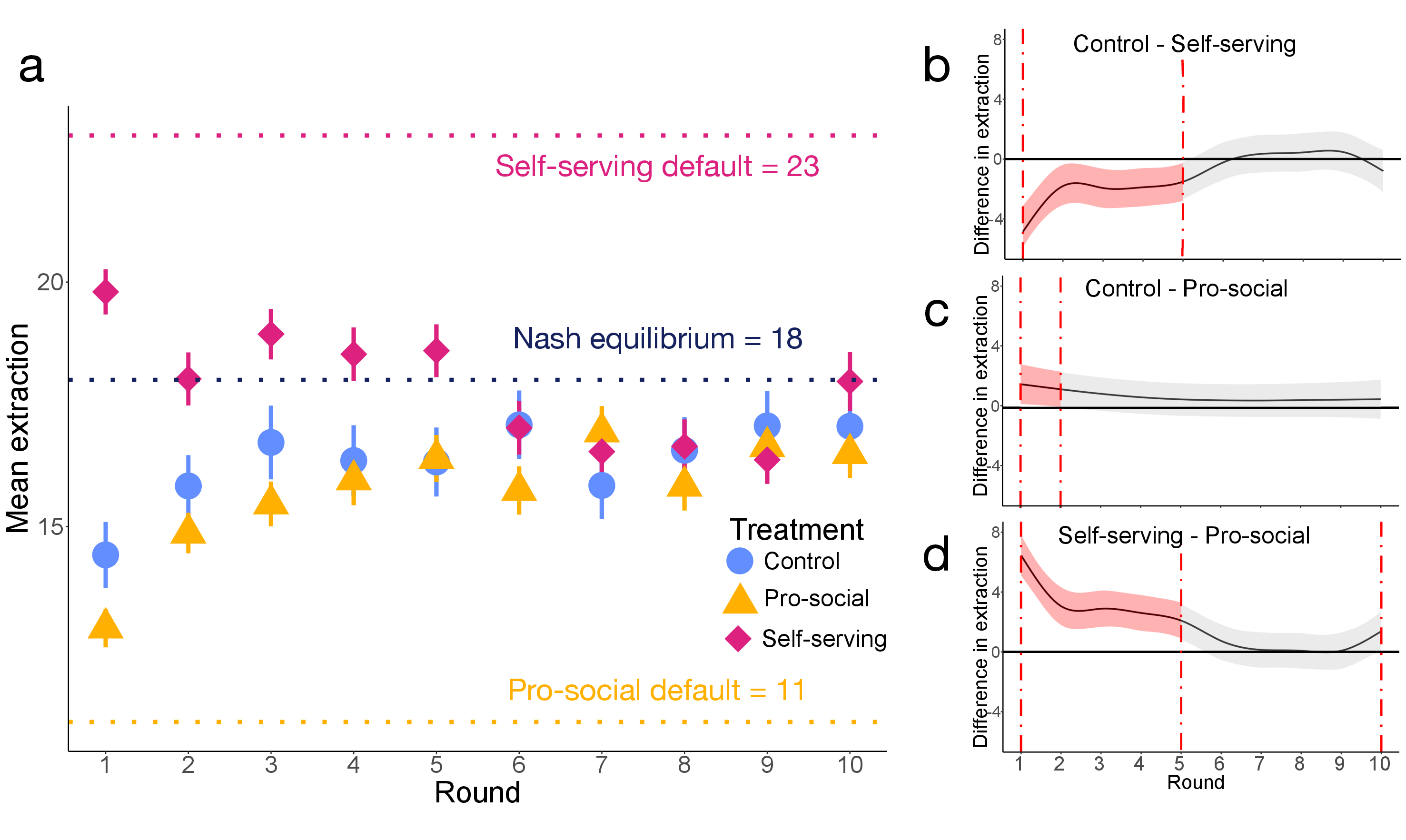}
\caption{\textbf{Mean extraction per treatment}. \textbf{A)} Mean extraction by round and by treatment: Pro-social $n = 156$, Self-serving $n = 156$ and Control $n = 100$. The blue dotted line represents the Nash Equilibrium of the game, while the dotted lines on both extremes represent the default values presented to participants. Vertical lines represent $95\%$  confidence interval. \textbf{(B, C and D)} Estimated difference of mean extraction between treatments the red is the estimated difference, as given by the mixed-effect model, while the vertical lines represent the round where this significant difference can be found. The colour shaded areas represent the $95\%$ confidence interval from the model.}\label{fig:line_treatments}
\end{figure}

In the experiment flow (see Figure \ref{fig:experiment_flow}), participants were randomly assigned to one of three treatments in the third stage (see Methods for details): In the \emph{Self-serving} treatment ($n = 156$ participants,  39 groups), the default extraction shown to participants in the first five rounds corresponds to an individually beneficial extraction (i.e. extraction value $x=23$). This default could potentially yield a high individual payoff, but if collectively chosen, results in all participants obtaining zero payoff. In the \emph{Pro-social} treatment ($n = 156$ participants,  39 groups), the default was set to an extraction value beneficial for resource and society (i.e. extraction value $x=11$). This choice is a mutually optimal one, which, if selected by everyone, leads to an individual payoff of $132$ ECU and maximises collective surplus. Lastly, in the \emph{Control} treatment ($n = 100$ participants,  25 groups), no default value was shown, and thus each participant had to select how much to consume in each round. In CPRD theory, rational, self-interested participants extract the symmetric Nash equilibrium ($x = 18$, depicted as the blue dashed line in Figure \ref{fig:line_treatments}), leading to resource overexploitation and a suboptimal equilibrium \cite{falk_appropriating_2001}.

In the first five rounds of the CPRD (see Figure \ref{fig:experiment_flow}), where participants of the Self-serving and Pro-social treatments were shown a default value, the mean extraction did not start exactly from the default option but was effectively nudged to more or less extraction levels depending on the default value presented to them, as shown in Figure \ref{fig:line_treatments}A. The pink markers in Figure \ref{fig:line_treatments}A show that participants in the Self-serving treatment extracted more ($\overline{x} = 18.7756$, 95\% CI = $[18.3171, 19.2342]$) in the first rounds than those in the control ($\overline{x} = 15.9280$, 95\% CI = $[15.3137, 16.5423]$) and Pro-social ($\overline{x} = 15.1192$, 95\% CI = $[14.7122, 15.5263]$) treatments. An analysis of variance (ANOVA) shows for the first five rounds that the average extraction over time is significantly different between treatments ($F(2) = 9.7317$, p-value$ = <0.0001$). The Self-serving default nudged participants towards the Nash equilibrium, particularly in rounds 3 to 5, with a mean extraction of $\overline{x} = 18.6859$ and a 95\% CI of $[18.0816, 19.2902]$, not significantly different from the Nash equilibrium (Wilcoxon signed-rank test, V $= 24780$, $p = 0.1734$).

As long as the default value was shown, the participants in the Self-serving treatment extracted significantly more than the in control treatment ($F(1) = 54.872$, p-value$ = <0.0001$). For the Pro-social treatment, this difference extended until round 3, but thereafter the difference was no longer significant ($F(1) = 1.0671$, p-value$ = 0.3017$). 

To visualise these differences, Figure \ref{fig:line_treatments}B, C and D show the estimated differences in the extraction over rounds using a mixed-effects model (see Methods for more details on this model).  This model can identify the time periods where the extraction is significantly different from zero. The area in pink between the red vertical lines represents the area where differences between treatments are significant at the 5\% level. The fit between the model and the experimental data can be seen in Figure \ref{fig:line_treatment_fit} in the Appendix.

Figure \ref{fig:line_treatments}B shows that the participants extracted less (below the zero line) in the control treatment than in the Self-serving treatment in the first five rounds, i.e. where the default value was shown. Figure \ref{fig:line_treatments}C shows how participants extracted more (above the zero line) in rounds 1 and 2, and then the mean extraction remained the same on average in the remaining rounds. Figure \ref{fig:line_treatments}D shows the mean difference in extraction between the Pro-social and Self-serving treatments. 

In this regard, the default effect showed an asymmetrical duration: the effect of showing a selfish default value lasted for longer, compared to showing a pro-social default value. Indeed, if the effect was symmetrical, participants in the Pro-social treatment would extract significantly less (compared to showing no default) for the same duration as they did in the Self-serving treatment. The selfish default was more effective to anchor participants to over-extract than the pro-social default was to nudge participants toward sustainable levels of extraction.

\subsection{The effect of all defaults disappear as soon as they are lifted}

After round 6, where participants must choose an extraction without a default as in the Control treatment, the participants in the three treatments had similar mean extractions, as shown in the gray shaded areas of Figures \ref{fig:line_treatments}B, C and D.  Actually, in the last five rounds, the mean extractions of the Self-serving ($\overline{x} = 16.9051$, 95\% CI = $[16.4246, 17.3857]$) and Pro-social ($\overline{x} = 16.3256$, 95\% CI = $[15.8844, 16.7668]$) treatments are very similar to the mean extraction in the Control treatment for all rounds ($\overline{x} = 16.7180$, 95\% CI = $[16.1161, 17.3199]$). An analysis of variance (ANOVA) further confirms this, as extractions are not significantly different between the three treatments after round 6 ($F(2) = 1.5333$, p-value$ = 0.2161$). Note that the mean extraction in all treatments is closer to the symmetric Nash equilibrium (18 tokens, shown in the blue dotted line in Figure \ref{fig:line_treatments}A) than to the default values provided in the treatments. Therefore, the effect of defaults fades away as soon as the defaults are lifted, which contradicts the ``stickiness'' hypothesis.

\subsection{Cooperatives can be nudged to extract more and selfish participants to extract less.}
\label{sec:results_2}

We hypothesised that extractions are affected not only by the defaults but also by personal preferences (which were measured in Task 1 and 2) and that there could be interesting interactions between the default value and both SVO and Risk preferences.

The results of Tasks 1 and 2 allows us to study the personal preferences of the participants and to link them with the behaviour in the third task. Figure \ref{fig:svo_gamble} in the Appendix presents the distributions of SVO, measuring social preferences, and gambles choices, measuring risk preferences. Most participants across all treatments fall under the ``cooperative'' spectrum ($n = 275, \approx 67\%, 22.45^{\circ} < SVO^\circ < 57.15^\circ$) and another share under the ``individualistic'' ($n = 132, \approx 32\%, 12.04^\circ < SVO^\circ < 22.45^\circ$) trait in this task. Three participants can be classified as ``altruistic'' ($\approx 0.7\%, SVO^\circ > 57.15^\circ$) and two as ``competitive'' ($\approx 0.4\%, SVO^\circ < 12.04^\circ$). Moreover, 283 participants (69\%) picked the three least risky gambles (the lower, the safest) in the Risk Assessment task. Figures \ref{fig:svo_gamble} and \ref{fig:svo_gamble_extraction} in the Appendix show the distribution of extraction by SVO scores and gamble choices.
 
To link the individual data from the first two tasks with the data of Task 3, we fitted a mixed-effects model (see Methods section for the details on the fitting). The results of the regression model are detailed in Table \ref{tab:mixed_ef} in the Appendix, there is a significant interaction term between SVO score and the treatments, which means that the extraction participants made depended on both the default value presented and their SVO score.

\begin{figure}
    \centering
    \includegraphics[width=1\linewidth]{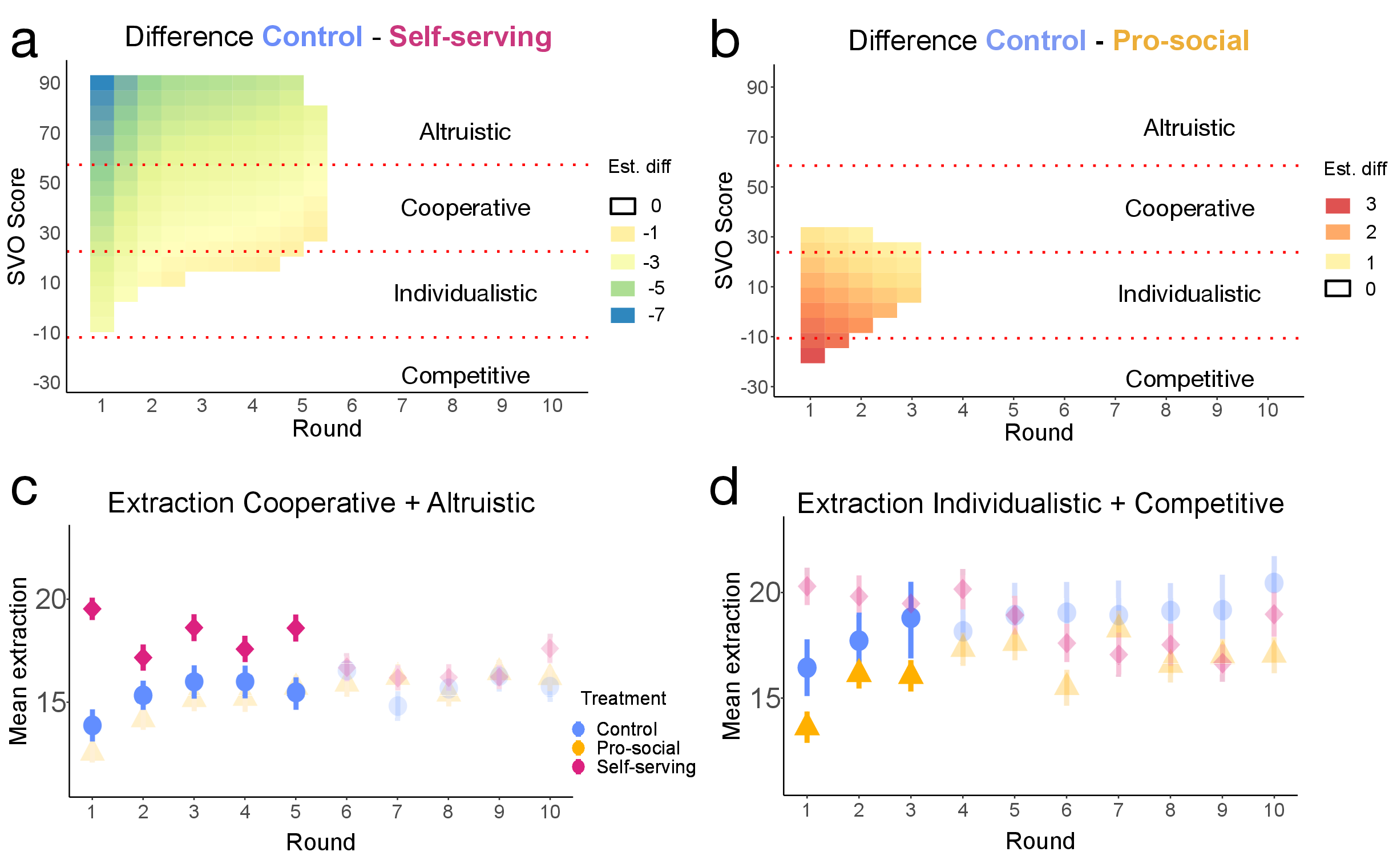}
    \caption{\textbf{Difference in extraction depending on participants' SVO over time}. \textbf{A and B:} Estimated difference given by the mixed-effects model, where non-white parts show a significant difference between two treatments. The horizontal lines show the SVO categories for reference and the colours the difference in extraction. Negative (or positive) differences show a larger (or smaller) extraction than the results in the Control treatment. \textbf{C and D:} Mean extraction over time given by the experimental data, where the most pro-social categories of SVO (cooperative and altruist) are grouped in panel \textbf{C} and the most selfish categories of SVO (individualistic and competitive) are grouped in panel \textbf{D}. The highlighted points in both panels are the ones given by the model in panels A and B. Vertical bars show the 95\% confidence interval.}
    \label{fig:svo_diff}
\end{figure}

The model also reveals a difference in extraction based on SVO and CPRD round, which we visualise in the form of a heatmap in Figure \ref{fig:svo_diff}. This figure shows the extraction difference between two treatments, and where a significant difference was found. In panel \textbf{A}, the difference shown is negative, meaning that the participants in the Self-serving treatment extracted more than in the Control treatment, for the first five rounds and mostly on the upper bounds of the SVO spectrum, i.e. where Cooperative and Altruistic preferences reside. This means that the Self-serving default presented led cooperative and altruistic individuals to extract more on average ($F(1) = 44.662$, p-value $<0.0001$) than what they would do in the Control, as also shown in Figure \ref{fig:svo_diff}C. 
In panel \textbf{B}, where the Pro-social treatment is compared to the Control, this difference is mostly in the first three rounds for the lower bounds of the SVO spectrum, i.e. Individualistic and Competitive participants. Similar to the previous finding, the Pro-social default makes selfish participants extract less, with respect to the Control treatment ($F(1) = 7.6922$, p-value$=0.0059$). This difference can be seen in Figure \ref{fig:svo_diff}D where the highlighted markers show the mean extraction in treatments Control and Pro-social.

Regarding risk preferences, we classified the subjects who picked gambles 1 to 3 as "risk-averse" and those who picked gambles 4 to 6 as "risk-seekers" (see Methods for details about the gambles). We found that in the Self-serving treatment, risk-seekers extracted less on average ($\overline{x} = 17.8809$, 95\% CI = $[16.9825, 18.7792]$) than risk-averse subjects ($\overline{x} = 19.1615$, 95\% CI = $[18.6331, 19.6899]$), and this difference is significant (KS-statistic $= 0.1181$, p-value $ = 0.0001$). Conversely, risk-seeking subjects extracted more in the Pro-social treatment ($\overline{x} = 16.3915$, 95\% CI = $[15.5705, 17.2124]$) than risk-averse ($\overline{x} = 14.5706$, 95\% CI = $[14.1138, 15.0275]$), and the difference between these two means is significant (KS Statistic $= 0.1450$, p-value$ = 0.001$), see Figure \ref{fig:gamble_default}. With this finding, we could observe how the risk-averse participants conform with the proposed default more often, resulting in more average extraction in the Self-serving treatment and less extraction in the Pro-social, with respect to their risk-seeking counterparts. 

\newpage

\section{Discussion}\label{sec:discussion}

Our experiments show that defaults impact over-extraction in a CPRD. In our case, we found an asymmetrical persistence of the influence of the default, with respect to the Control treatment. Specifically, participants in the Self-serving treatment extracted more on average over five rounds, while those in the Pro-social group extracted less only in the initial two rounds. Notably, our findings indicate that this effect persists even when players are repeatedly prompted with the same decision, as shown in previous research \cite{van_rookhuijzen_effects_2021}. 

Previous literature has shown that individuals often underestimate the influence of default choices on their decisions \cite{zlatev_default_2017}. Echoing this, in our study, nearly half of the participants (161 out of 312) across both Self-serving and Pro-social treatments reported feeling unaffected by the default values presented. Nonetheless, our analysis showed the default value significantly influenced average extractions in each treatment. However, our analysis showed that the extractions of those who self-assessed as ``unaffected'' are significantly different from the mean extraction in the Control (lower for the Pro-social treatment and higher for the Self-serving treatment) indicating a disconnect between perceived and actual influence of defaults.


In the CPRD, participants are incentivised to extract more than their peers. Although their extractions decreased with respect to the control, in the first five rounds of the Pro-social treatment, participants still extracted more than the presented default. In contrast, in the Self-serving default, participants extracted significantly more than in the control, and converged towards an average extraction which coincides with the Nash equilibrium of the game. Additionally, in both Pro-social and Self-serving treatments, there is a symmetrical trend of converging towards the average extraction observed in the control, once the default nudge is removed. However, the time it takes for the population of each treatment to converge is asymmetrical, that is, the effect of the selfish default lasted longer than the effect of the pro-social one. 

Nevertheless, our results show that if the participants play the CPRD game with high values by default (Self-serving treatment), it is expected for them to extract significantly more than showing a low default (Pro-social treatment). Since our experiment intended to reflect real scenarios of resource consumption, this result is highly important. For example, in contexts such as energy consumption, individuals often face the choice of selecting a consumption plan, with one option frequently set as the default. This design decision can ultimately influence both individual payoffs and resource consumption patterns.

In rounds 6-10, participants had to pick their extraction in every round without a default, regardless of the treatment. One of our hypotheses was that we might observe a persistence effect of the default, such as participants continuing to extract at a higher rate if they were initially shown a selfish default in the first five rounds. However, our results did not support this hypothesis. We found no difference in mean extraction once the default was lifted. Participants in all treatments extracted around 16 tokens per round in all treatments. This quantity is closer to the symmetrical Nash equilibrium \cite{ostrom_rules_1994} than to the default values provided in the treatments, which is still inefficient from the resource management point of view \cite{saijo_common-pool_2017}.

Another important finding relates to how individual preferences (as measured by the SVO) determine resource allocation. We wanted to understand the effect of the (mis)alignment of the default value and the participants' SVO: i.e. what happens if a cooperative person is confronted with a selfish default and vice-versa. Previous research suggests that individuals with higher SVO scores typically exhibit more pro-social behaviour, while those with lower scores tend toward more self-interested behaviour. \cite{joireman_interdependence_1997, fleis_social_2019} However, our results show that cooperative individuals extracted more when shown a selfish default than in the absence of default. Moreover, defaults worked as a nudge to make selfish participants extract less when confronted with a low default as in the Pro-social treatment, confirming the use of default values as a valuable tool in fostering sustainable behaviour.


The results presented in this paper shed light on some of the factors - besides economic incentive - that play a role in how we make decisions in common pool resource dilemmas. We show that participants can be nudged into a certain behaviour by pre-setting an extraction value. This has implications on the way we consume finite shared resources and how we interact in such a social context. We found that selfish participants (as measured by the SVO) were induced to extract less by a default, suggesting a sustainable course of action. In the case of systems where a default value has to be enforced by design, setting up the "best practices" as defaults, may be an effective strategy to mitigate over-consumption. On the other hand, setting the "wrong" default can have even stronger (negative) consequences, and spoil the decisions of previously pro-social participants \cite{kobis_bad_2021}. While good defaults can nudge otherwise selfish people, bad defaults have the potential to do the inverse to a greater extent. 

This last point is critical for policy-making, for example with laws aimed at mitigating dark patterns in digital services and markets \cite{european_parlament_regulation_2022}, enhancing data privacy \cite{european_parlament_regulation_2016}, and improving products' environmental performance \cite{european_parlament_directive_2009} through the use of defaults. Policies should adopt evidence-based approaches to ensure defaults align with individual and societal welfare, emphasising ethical considerations and autonomy preservation of the individuals.

\section{Limitations}

Our experiment uses an abstract game as a proxy for real-world resource consumption. To improve the generalisation of our results, we avoided using loaded language or contextual instructions. While this allows for a greater control and interpretability of participants' behaviour, this also sidesteps important dimensions that are present in many (politicised) real-world settings, such as emotional involvement in the preservation of the environment, notions of fairness or the idea of global responsibility. Thus, future use of our findings may require further testing to account for contextual specificity of the desired application setting.

Moreover, our study did not account for the long-term behavioural changes of participants, which suggests the need for longitudinal studies to observe the durability of the observed behaviours in the long term.

\section{Conclusions}

Our experiment reveals that exposing participants to a default extraction value in a CPRD significantly affects their extraction patterns. Specifically, we found that participants in the Self-serving treatment extracted significantly more resources than those not exposed to any default values. Additionally, its impact varies based on the participant's SVO. Cooperative individuals exhibited selfish behaviour by extracting more when faced with a Self-serving default, while, notably, selfish individuals extracted less when confronted with a Pro-social default. Furthermore, the default effect is non-persistent, and its impact on participants' decisions diminishes rapidly once the default is removed.

Our findings have substantial implications for designing systems and policies aimed at the sustainable management of common resources.
They highlight the importance of accounting for individual heterogeneity and demonstrate the potential of simple, low-cost interventions like defaults in nudging groups towards more socially beneficial outcomes.

\section{Methods}\label{sec:methods}

\subsection{Study design}\label{sec:methods_study_design}

The data used in this paper was collected through a behavioural experiment, performed on the online platform Prolific (\href{(www.prolific.com)}{www.prolific.com}). Participants had to complete three tasks (see Figure \ref{fig:experiment_flow}). Task 1 is the Social Value Orientation test (SVO) from Murphy et al. \cite{murphy_measuring_2011}, and task 2 is the Risk-attitude elicitation task developed by Eckel and Grossman \cite{eckel_sex_2002} with the values used by Dave and Eckel \cite{dave_eliciting_2010}. In task 3, participants play a CPRD in a group of 4. We designed the experiment so we can link extraction behaviour in the CPRD in task 3 with participants' social preferences and  risk attitudes elicited in tasks 1 and 2. Additionally,  our experimental design, hypotheses, analyses and sample size were pre-registered in OSF: \href{https://osf.io/jg2sa}{https://osf.io/jg2sa}



\subsubsection{Common Pool Resource Dilemma}\label{sec:methods_CPR}

In the main task, task 3, participants played a version of the Common Pool Resource Dilemma (CPRD). This dilemma captures the tension that emerges in a group exploiting a common finite resource. The group payoff is determined by the sum of extractions of the group members: over-extraction  depletes the resource, leading to a payoff of zero to everyone, while a sustainable management of the resource yields the highest collective payoff. Individual payoff is proportional to individual extraction: one gets a higher share of the group payoff if one extracts a bigger share of the group extraction.  Hence, narrow self-interest dictates to appropriate the resource as much as possible, however, if everyone does so, depletion ensues (tragedy of the commons), which is the worst collective outcome. 

We propose a variant of the game where participants have to extract a minimum of 1 and maximum 30 units (or tokens, as they are called in the experiment) from the resource, which participants can select from a drop-down menu in each round. The extraction variable is discrete, that is, participants were able to extract any natural number in the interval $[1..30]$. Participants can aim to maximise their individual payoffs by extracting more from the resource, or they can aim to have a group maximum by coordinating their strategies to do so. 

The amount extracted by each player $i$ from the resource is $x_i$, and the amount extracted by the group is $X = \sum_{i=1}^{N} x_i$. Extraction of the resource earns each player $a$ times  every unit extracted personally, minus $b$ for every unit extracted by the group regardless of who extracts it. We adjusted the parameters used by Walker et al. \cite{walker_rent_1990} to obtain similar payoffs among the different tasks ($a = 2.3$ and $b = 0.025$). The payoff of extracting $x_i$ and the group extraction $X$ from the CPRD is then:

\begin{equation}
    \pi_i =  (x_i/X)(aX - bX^2)	
\end{equation}

\begin{figure}
    \centering
    \includegraphics[width=1\linewidth]{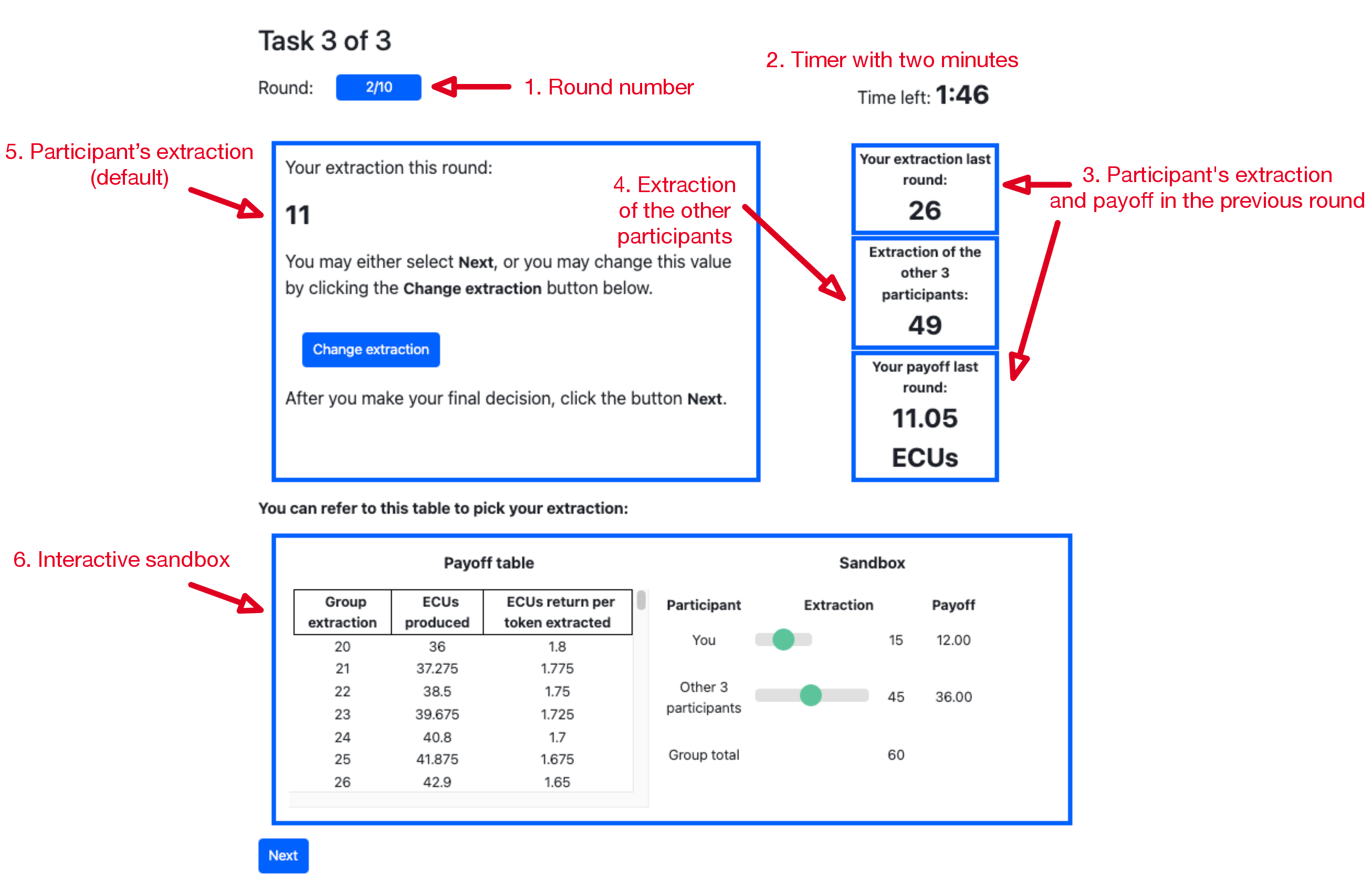}
    \caption{\textbf{Screenshot of the third task}. In this task, participants had to choose their desired extraction for ten rounds. At the top, the platform showed the round number and the time left to make their decision. The participants had their previous extraction and the extraction of the other members of the group, and their payoff, as shown on the left side of the figure. At the bottom, participants had a sandbox at their disposal to calculate their potential earnings depending on theirs and others' extraction.}
    \label{fig:screenshot}
\end{figure}

Participants have the following information at all times (see Figure \ref{fig:screenshot}):

\begin{enumerate}
    \item The round number.
    \item A two-minutes timer.
    \item The participant's extraction and payoff in the previous round
    \item The total group extraction in the previous round
    \item The participant's extraction for the round (this is where the default was shown)
    \item An interactive sandbox where they can simulate their and others' payoffs and a table with all the possible group extractions, and the ECU's produced.
\end{enumerate}

Additionally, the participants knew that they are interacting in groups of $4$, that games is played over $10$ rounds, and that there is a maximum of $2$ minutes to make a decision. Participants were instructed to make their decisions within this time, otherwise they will be considered as dropouts from the experiment. Also, after every round, the participants know how much the other three participants extracted in the previous round, and they are notified if a co-player dropped out of the experiment, in which case they continued playing in a smaller group, and got paid. Dropouts, totalling 21 instances ($2.8\%$ of the total participants) due to \textit{Timeout} or \textit{Lost focus} (see Table \ref{tab:status}), led to the exclusion of data from those completing the ten rounds. Table \ref{tab:dropouts} indicates that previous tasks' results, demographic information, or approval numbers do not predict individual drop-out, discarding selective attrition concerns.

All participants had to complete a comprehension test for task 3, in which they were given five attempts to get the five questions correct. If they could not complete this test, they were paid for the earnings in the previous two tasks, but they were excluded from playing the CPRD game. All the instructions and captions can be found in the supplementary documents. 

\subsubsection{Experimental treatments}\label{sec:methods_default_values}

The treatments consist in the introduction of default extractions. We implemented defaults by pre-setting a token extraction value. Participants had the possibility to override this value. We also include a Control treatment with no default values, i.e., participants always had to choose their own extraction. Both in the Control treatment and the last five rounds of the Pro-social and Self-serving treatments, this was enforced by making participants choose a value between 1 and 30, before proceeding to the next round. In this case, what they saw in the interface then was a drop-down menu, with two dashes (--) meaning that they had to pick an amount. We compare the effect of the manipulation done in the two default value treatments with this control. 

The values were chosen based on the social optimum and the individualistic extraction: 

\begin{itemize}
    \item Social optimum: $x_f = \frac{a}{2bn}$ where n is the group size.
    \item Individualistic extraction: $x_h = \frac{a}{bn}$
\end{itemize}

Given the parameters we chose ($a = 2.3$ and $b = 0.025$, see previous section), $x_f = 11.5$ and $x_h = 23$. We rounded $x_f$ to $11$ because participants can only pick an integer from the user interface. If in any given round, $x_f$ is collectively chosen, i.e. the group extraction is 44, the resource will yield its maximum. This also means that participants can get more if they stick with this extraction. However, players will be enticed to pick a higher extraction to get a higher share of the group extraction. When the group extraction reaches 92, or $x_h$ is chosen by everyone, the group and individual payoffs are zero. The Nash equilibrium in this game is defined as $x^* = \frac{a}{b(n+1)}$, in which, if all players are self-interested and rational, they will pick this reaching a suboptimal equilibrium \cite{saijo_common-pool_2017, bravo_agents_2011}, as shown in Figure \ref{fig:group_extraction_payoff} in the Appendix.

The group payoff depending on the total of extraction can be found in Figure \ref{fig:group_extraction_payoff} in the Appendix. The coloured dashed lines in the figure indicate the amount that is given back to the players if all members of a group selected a certain default value. 

Participants are subjected to one of the 3 treatments:

\begin{itemize}
    \item \textbf{Control - No default value, $n = 100$}: in this control treatment, participants play the CPRD without default extraction proposed for 10 rounds.
    \item \textbf{Pro-social treatment (Pro-social), $n = 156$}: in this treatment, participants are shown a default value representing the social optimum value $x_f$ with a label that reads: \textit{Your extraction this round:} $11$. At each round, each participant has the option to override this value, and choose another extraction for themselves. The wording is neutral, and no notion of “fairness” or “fair extraction” is communicated to the participants. 
    \item \textbf{Self-serving treatment (Self-serving), $n = 156$}: in this treatment, participants are shown a default value representing the individualistic value $x_h$ with  a label that reads: \textit{Your extraction this round:} $23$. At each round, each participant has the option to override this value, and choose another extraction for themselves. The wording is neutral, and no notion of “individualistic” or “selfish extraction” is communicated to the participants. 
\end{itemize}

To change the default extraction, participants have to click a button below the extraction and pick a value from a drop-down menu. To measure persistence, in both treatments (Pro-social and Self-serving) participants are subjected to the default value manipulation for 5 rounds, while in the subsequent 5 rounds, no default option is presented, i.e., participants have to manually pick the extraction they desire. Our experimental manipulation is ``between-subject'': participants could not take part in more than one treatment. 

The currency used in this experiment is ECU (Experimental Currency Unit). Each ECU is converted to U.K. Pounds (£) with an exchange rate of or 100 ECUs = £1, or 1 ECU =  £0.01. Participants were recruited using Prolific (\href{www.prolific.com}{www.prolific.com}). In Prolific, we recruited participants who indicated they were fluent in English, had a high approval rate ($+99\%)$ and had more than $20$ submissions on the site. These filters were applied to filter out possible dropouts. Participants received £3 as a fixed participation fee plus £6 bonus depending on the decisions they made in the game. The average total earnings were £5.05 for a duration of roughly 30 minutes.

We excluded participants who did not sign the Informed Consent Form, dropped out, failed the comprehension test, or did not act within the time limit for each round. Participants with group drop-outs are also excluded, even if they finished the experiment themselves. Additionally, participants who were excluded or dropped out from the experiment were only paid for the tasks they completed. This design choice aimed at motivating participants to finish all tasks, while rightfully paying them for their time and for the completed tasks. Therefore, the experimental data used in this paper are from participants in groups of four who completed all ten rounds, passed the comprehension test, and successfully submitted their work on Prolific.  

The raw experimental data, instructions with screenshots included, is attached to the Supplementary Information. All experiments described in this paper followed the guidelines and regulations of data protection and experiments with human participants, and they were approved by the Ethical Commission for Human Sciences at the Vrije Universiteit Brussel in Brussels, Belgium (ECHW\_361.02). Also, all participants who took part in the experiment signed an Informed Consent form for the use of the data collected in the experiment, including decisions and background information, including gender and age. Without signing this form, they could not proceed with the experiment.

\subsubsection{Personal preferences}\label{sec:methods_svo}

For the first task of the experiment, we assessed participants' social preferences using the Social Value Orientation measure (SVO) \cite{lange_pursuit_1999} and the Slider measure by Murphy et al. \cite{murphy_measuring_2011, murphy_social_2013} with a modification to include incentives (see Figure \ref{fig:svo_sample_decisions} for a sample of the decisions shown to participants). Participants made six allocation decisions. Participants were informed that they would be matched with another participant at the end of the experiment. In each match, the allocation decisions of one of the participants would be played out (described as Group A), with the other one being paid as a receiver (described as Group B). When decisions are made, one of the six choices was randomly selected as the final allocation between a participant from Group A and another from Group B. Participants in Group B will receive an allocation determined by a member of Group A. Their resulting SVO were represented as an angle from the origin, where a higher angle means greater cooperativeness, and lower angles indicated individualistic or competitive allocations. For an easier analysis, we grouped subjects with SVO angles $< 22.45^{\circ}$ as "Individualistic+Competitive" and SVO $>= 22.45^{\circ}$ as "Cooperative+Altruistic" (as done in other works \cite{roch_effects_1997, liebrand_value_1986}) in some sections of the document instead of using the continuous angle.

In the second task, we implemented a risk-attitude elicitation task as done by Eckel and Grossman \cite{eckel_sex_2002}, with the values from Dave et al. \cite{dave_eliciting_2010} Participants have to decide between six different gambles, from the least risky to the most risky, of an event happening with 50\% chance. A coin was flipped to pick which of the events will pay them and that was their payoff for their task. The resulting measure is a discrete range from 1 to 6 where 1 is the safest and 6 is the riskiest (see Figure \ref{fig:risk_sample_decisions} for a sample of the options shown to participants). 

\subsubsection{Mixed models}\label{sec:methods_mixed_models}

We fitted a generalized additive (mixed) regression model (GAM) for the panel data using the \texttt{mgcv} \emph{R} package for \emph{GAM} estimation \cite{wood_fast_2011, wood_mgcv_2023}. We used this to test the interactions of the variables over time and by treatment. The variables we want to study to understand the extraction behaviour were:

\begin{itemize}
    \item Treatment (or the default value presented, if any)
    \item Round number
    \item SVO score in task 1
    \item Gamble choice in task 2
    \item Player's extraction in the last round
    \item Others' extraction in the last round
\end{itemize}

Our aim is to test the hypothesis that the default value presented in the Pro-social and Self-serving treatments will have different effects in function of the social and risk preferences of different subjects. Specifically, depending on their SVO score in task 1 and gambling choice in task 2. The reason behind this is that we hypothesised, for example, that cooperative individuals will react to a high default differently compared with an individualistic person. Also, we assumed random intercepts for each individual as a random effect, that accounts for individual-level variability in our subject pool, not captured by the fixed effects we included. Hence, our \texttt{lme4} formula looked like this:

\begin{equation}
    x_{i,r} \sim T + T \times \text{SVO} + T \times r + T \times \text{gamble} + x_{i, r-1} + X_{-i,r-1} + (1 | i)
\end{equation}

Where $x_{i,r}$ is the extraction of the player $i$ in round $r$ where they were in treatment $T$, also $X_{-i,r-1}$ represents the extraction of the three other group members, and the last term $(1 | i)$ in the equation represents the random effects for player variability.

We also show the difference in extraction between two levels of a categorical variable, in this case we use this model to show the time window where the extraction between two treatments is significantly different according to the model. A significant time window is given by the estimated difference in extraction, where the confidence intervals of the prediction do not include zero. This is useful to show the evolution of the default manipulation and how it affects individuals with different SVO scores. 


\begin{thebibliography}{10}
\expandafter\ifx\csname url\endcsname\relax
  \def\url#1{\burl{#1}}\fi
\expandafter\ifx\csname urlprefix\endcsname\relax\def\urlprefix{URL }\fi
\providecommand{\bibinfo}[2]{#2}
\providecommand{\eprint}[2][]{\url{#2}}
\providecommand{\doi}[1]{\url{https://doi.org/#1}}
\bibcommenthead

\bibitem{hardin_tragedy_1968}
\bibinfo{author}{Hardin, G.}
\newblock \bibinfo{title}{The {Tragedy} of the {Commons}}.
\newblock \emph{\bibinfo{journal}{Science}} \textbf{\bibinfo{volume}{162}}, \bibinfo{pages}{1243--1248} (\bibinfo{year}{1968}).
\newblock \urlprefix\url{https://www.jstor.org/stable/1724745}.
\newblock \bibinfo{note}{Publisher: American Association for the Advancement of Science}.

\bibitem{ostrom_rules_1994}
\bibinfo{author}{Ostrom, E.}, \bibinfo{author}{Gardner, R.}, \bibinfo{author}{Walker, J.}, \bibinfo{author}{Walker, J.~M.} \& \bibinfo{author}{Walker, J.}
\newblock \emph{\bibinfo{title}{Rules, {Games}, and {Common}-pool {Resources}}}  (\bibinfo{publisher}{University of Michigan Press}, \bibinfo{year}{1994}).
\newblock \bibinfo{note}{Google-Books-ID: DgmLa8gPo4gC}.

\bibitem{abrahamse_review_2005}
\bibinfo{author}{Abrahamse, W.}, \bibinfo{author}{Steg, L.}, \bibinfo{author}{Vlek, C.} \& \bibinfo{author}{Rothengatter, T.}
\newblock \bibinfo{title}{A review of intervention studies aimed at household energy conservation}.
\newblock \emph{\bibinfo{journal}{Journal of Environmental Psychology}} \textbf{\bibinfo{volume}{25}}, \bibinfo{pages}{273--291} (\bibinfo{year}{2005}).
\newblock \urlprefix\url{https://www.sciencedirect.com/science/article/pii/S027249440500054X}.

\bibitem{momsen_intention_2014}
\bibinfo{author}{Momsen, K.} \& \bibinfo{author}{Stoerk, T.}
\newblock \bibinfo{title}{From intention to action: {Can} nudges help consumers to choose renewable energy?}
\newblock \emph{\bibinfo{journal}{Energy Policy}} \textbf{\bibinfo{volume}{74}}, \bibinfo{pages}{376--382} (\bibinfo{year}{2014}).
\newblock \urlprefix\url{https://www.sciencedirect.com/science/article/pii/S0301421514004121}.

\bibitem{setlhaolo_optimal_2014}
\bibinfo{author}{Setlhaolo, D.}, \bibinfo{author}{Xia, X.} \& \bibinfo{author}{Zhang, J.}
\newblock \bibinfo{title}{Optimal scheduling of household appliances for demand response}.
\newblock \emph{\bibinfo{journal}{Electric Power Systems Research}} \textbf{\bibinfo{volume}{116}}, \bibinfo{pages}{24--28} (\bibinfo{year}{2014}).
\newblock \urlprefix\url{https://www.sciencedirect.com/science/article/pii/S0378779614001527}.

\bibitem{jachimowicz_when_2019}
\bibinfo{author}{Jachimowicz, J.~M.}, \bibinfo{author}{Duncan, S.}, \bibinfo{author}{Weber, E.~U.} \& \bibinfo{author}{Johnson, E.~J.}
\newblock \bibinfo{title}{When and why defaults influence decisions: a meta-analysis of default effects}.
\newblock \emph{\bibinfo{journal}{Behavioural Public Policy}} \textbf{\bibinfo{volume}{3}}, \bibinfo{pages}{159--186} (\bibinfo{year}{2019}).
\newblock \urlprefix\url{https://www.cambridge.org/core/journals/behavioural-public-policy/article/when-and-why-defaults-influence-decisions-a-metaanalysis-of-default-effects/67AF6972CFB52698A60B6BD94B70C2C0}.
\newblock \bibinfo{note}{Publisher: Cambridge University Press}.

\bibitem{fischbacher_social_2010}
\bibinfo{author}{Fischbacher, U.} \& \bibinfo{author}{Gächter, S.}
\newblock \bibinfo{title}{Social {Preferences}, {Beliefs}, and the {Dynamics} of {Free} {Riding} in {Public} {Goods} {Experiments}}.
\newblock \emph{\bibinfo{journal}{American Economic Review}} \textbf{\bibinfo{volume}{100}}, \bibinfo{pages}{541--556} (\bibinfo{year}{2010}).
\newblock \urlprefix\url{https://www.aeaweb.org/articles?id=10.1257/aer.100.1.541}.

\bibitem{behlen_defaults_2023}
\bibinfo{author}{Behlen, L.}, \bibinfo{author}{Himmler, O.} \& \bibinfo{author}{Jäckle, R.}
\newblock \bibinfo{title}{Defaults and effortful tasks}.
\newblock \emph{\bibinfo{journal}{Experimental Economics}}  (\bibinfo{year}{2023}).
\newblock \urlprefix\url{https://doi.org/10.1007/s10683-023-09808-8}.

\bibitem{taube_how_2019}
\bibinfo{author}{Taube, O.} \& \bibinfo{author}{Vetter, M.}
\newblock \bibinfo{title}{How green defaults promote environmentally friendly decisions: {Attitude}-conditional default acceptance but attitude-unconditional effects on actual choices}.
\newblock \emph{\bibinfo{journal}{Journal of Applied Social Psychology}} \textbf{\bibinfo{volume}{49}}, \bibinfo{pages}{721--732} (\bibinfo{year}{2019}).
\newblock \urlprefix\url{https://onlinelibrary.wiley.com/doi/abs/10.1111/jasp.12629}.
\newblock \bibinfo{note}{\_eprint: https://onlinelibrary.wiley.com/doi/pdf/10.1111/jasp.12629}.

\bibitem{liebe_large_2021}
\bibinfo{author}{Liebe, U.}, \bibinfo{author}{Gewinner, J.} \& \bibinfo{author}{Diekmann, A.}
\newblock \bibinfo{title}{Large and persistent effects of green energy defaults in the household and business sectors}.
\newblock \emph{\bibinfo{journal}{Nature Human Behaviour}} \textbf{\bibinfo{volume}{5}}, \bibinfo{pages}{576--585} (\bibinfo{year}{2021}).
\newblock \urlprefix\url{https://www.nature.com/articles/s41562-021-01070-3}.
\newblock \bibinfo{note}{Number: 5 Publisher: Nature Publishing Group}.

\bibitem{walker_rent_1990}
\bibinfo{author}{Walker, J.~M.}, \bibinfo{author}{Gardner, R.} \& \bibinfo{author}{Ostrom, E.}
\newblock \bibinfo{title}{Rent dissipation in a limited-access common-pool resource: {Experimental} evidence}.
\newblock \emph{\bibinfo{journal}{Journal of Environmental Economics and Management}} \textbf{\bibinfo{volume}{19}}, \bibinfo{pages}{203--211} (\bibinfo{year}{1990}).
\newblock \urlprefix\url{https://www.sciencedirect.com/science/article/pii/009506969090069B}.

\bibitem{alpizar_effect_2010}
\bibinfo{author}{Alpizar, F.}, \bibinfo{author}{Carlsson, F.} \& \bibinfo{author}{Naranjo, M.}
\newblock \bibinfo{title}{The {Effect} of {Risk}, {Ambiguity} and {Coordination} on {Farmers}’ {Adaptation} to {Climate} {Change}: {A} {Framed} {Field} {Experiment}}.
\newblock \bibinfo{type}{{SSRN} {Scholarly} {Paper}} \bibinfo{number}{ID 1646882}, \bibinfo{institution}{Social Science Research Network}, \bibinfo{address}{Rochester, NY} (\bibinfo{year}{2010}).
\newblock \urlprefix\url{https://papers.ssrn.com/abstract=1646882}.

\bibitem{bernedo_del_carpio_community-based_2021}
\bibinfo{author}{Bernedo Del~Carpio, M.}, \bibinfo{author}{Alpizar, F.} \& \bibinfo{author}{Ferraro, P.~J.}
\newblock \bibinfo{title}{Community-based monitoring to facilitate water management by local institutions in {Costa} {Rica}}.
\newblock \emph{\bibinfo{journal}{Proceedings of the National Academy of Sciences}} \textbf{\bibinfo{volume}{118}}, \bibinfo{pages}{e2015177118} (\bibinfo{year}{2021}).
\newblock \urlprefix\url{https://www.pnas.org/doi/10.1073/pnas.2015177118}.
\newblock \bibinfo{note}{Publisher: Proceedings of the National Academy of Sciences}.

\bibitem{ostrom_beyond_2010}
\bibinfo{author}{Ostrom, E.}
\newblock \bibinfo{title}{Beyond {Markets} and {States}: {Polycentric} {Governance} of {Complex} {Economic} {Systems}}.
\newblock \emph{\bibinfo{journal}{The American Economic Review}} \textbf{\bibinfo{volume}{100}}, \bibinfo{pages}{641--672} (\bibinfo{year}{2010}).
\newblock \urlprefix\url{https://www.jstor.org/stable/27871226}.
\newblock \bibinfo{note}{Publisher: American Economic Association}.

\bibitem{schill_sustaining_2023}
\bibinfo{author}{Schill, C.} \& \bibinfo{author}{Rocha, J.~C.}
\newblock \bibinfo{title}{Sustaining local commons in the face of uncertain ecological thresholds: {Evidence} from a framed field experiment with {Colombian} small-scale fishers}.
\newblock \emph{\bibinfo{journal}{Ecological Economics}} \textbf{\bibinfo{volume}{207}}, \bibinfo{pages}{107695} (\bibinfo{year}{2023}).
\newblock \urlprefix\url{https://www.sciencedirect.com/science/article/pii/S0921800922003561}.

\bibitem{falk_appropriating_2001}
\bibinfo{author}{Falk, A.}, \bibinfo{author}{Fehr, E.} \& \bibinfo{author}{Fischbacher, U.}
\newblock \bibinfo{title}{Appropriating the {Commons} - a {Theoretical} {Explanation}}.
\newblock \bibinfo{type}{{SSRN} {Scholarly} {Paper}} \bibinfo{number}{ID 287462}, \bibinfo{institution}{Social Science Research Network}, \bibinfo{address}{Rochester, NY} (\bibinfo{year}{2001}).
\newblock \urlprefix\url{https://papers.ssrn.com/abstract=287462}.

\bibitem{mertens_effectiveness_2022}
\bibinfo{author}{Mertens, S.}, \bibinfo{author}{Herberz, M.}, \bibinfo{author}{Hahnel, U. J.~J.} \& \bibinfo{author}{Brosch, T.}
\newblock \bibinfo{title}{The effectiveness of nudging: {A} meta-analysis of choice architecture interventions across behavioral domains}.
\newblock \emph{\bibinfo{journal}{Proceedings of the National Academy of Sciences of the United States of America}} \textbf{\bibinfo{volume}{119}}, \bibinfo{pages}{e2107346118} (\bibinfo{year}{2022}).

\bibitem{fosgaard_nudge_2015}
\bibinfo{author}{Fosgaard, T.~R.} \& \bibinfo{author}{Piovesan, M.}
\newblock \bibinfo{title}{Nudge for (the {Public}) {Good}: {How} {Defaults} {Can} {Affect} {Cooperation}}.
\newblock \emph{\bibinfo{journal}{PLOS ONE}} \textbf{\bibinfo{volume}{10}}, \bibinfo{pages}{e0145488} (\bibinfo{year}{2015}).
\newblock \urlprefix\url{https://journals.plos.org/plosone/article?id=10.1371/journal.pone.0145488}.
\newblock \bibinfo{note}{Publisher: Public Library of Science}.

\bibitem{bynum_passive_2016}
\bibinfo{author}{Bynum, A.}, \bibinfo{author}{Kline, R.} \& \bibinfo{author}{Smirnov, O.}
\newblock \bibinfo{title}{Passive non-participation versus strategic defection in a collective risk social dilemma}.
\newblock \emph{\bibinfo{journal}{Journal of Theoretical Politics}} \textbf{\bibinfo{volume}{28}}, \bibinfo{pages}{138--158} (\bibinfo{year}{2016}).
\newblock \urlprefix\url{https://doi.org/10.1177/0951629815586880}.
\newblock \bibinfo{note}{Publisher: SAGE Publications Ltd}.

\bibitem{ferguson_when_2020}
\bibinfo{author}{Ferguson, E.}, \bibinfo{author}{Shichman, R.} \& \bibinfo{author}{Tan, J. H.~W.}
\newblock \bibinfo{title}{When {Lone} {Wolf} {Defectors} {Undermine} the {Power} of the {Opt}-{Out} {Default}}.
\newblock \emph{\bibinfo{journal}{Scientific Reports}} \textbf{\bibinfo{volume}{10}}, \bibinfo{pages}{8973} (\bibinfo{year}{2020}).
\newblock \urlprefix\url{https://www.nature.com/articles/s41598-020-65163-1}.
\newblock \bibinfo{note}{Number: 1 Publisher: Nature Publishing Group}.

\bibitem{pichert_green_2008}
\bibinfo{author}{Pichert, D.} \& \bibinfo{author}{Katsikopoulos, K.~V.}
\newblock \bibinfo{title}{Green defaults: {Information} presentation and pro-environmental behaviour}.
\newblock \emph{\bibinfo{journal}{Journal of Environmental Psychology}} \textbf{\bibinfo{volume}{28}}, \bibinfo{pages}{63--73} (\bibinfo{year}{2008}).
\newblock \urlprefix\url{https://www.sciencedirect.com/science/article/pii/S0272494407000758}.

\bibitem{truelove_positive_2014}
\bibinfo{author}{Truelove, H.~B.}, \bibinfo{author}{Carrico, A.~R.}, \bibinfo{author}{Weber, E.~U.}, \bibinfo{author}{Raimi, K.~T.} \& \bibinfo{author}{Vandenbergh, M.~P.}
\newblock \bibinfo{title}{Positive and negative spillover of pro-environmental behavior: {An} integrative review and theoretical framework}.
\newblock \emph{\bibinfo{journal}{Global Environmental Change}} \textbf{\bibinfo{volume}{29}}, \bibinfo{pages}{127--138} (\bibinfo{year}{2014}).
\newblock \urlprefix\url{https://linkinghub.elsevier.com/retrieve/pii/S0959378014001599}.

\bibitem{cappelletti_are_2014}
\bibinfo{author}{Cappelletti, D.}, \bibinfo{author}{Mittone, L.} \& \bibinfo{author}{Ploner, M.}
\newblock \bibinfo{title}{Are default contributions sticky? {An} experimental analysis of defaults in public goods provision}.
\newblock \emph{\bibinfo{journal}{Journal of Economic Behavior \& Organization}} \textbf{\bibinfo{volume}{108}}, \bibinfo{pages}{331--342} (\bibinfo{year}{2014}).
\newblock \urlprefix\url{https://www.sciencedirect.com/science/article/pii/S0167268114000080}.

\bibitem{ghesla_nudge_2019}
\bibinfo{author}{Ghesla, C.}, \bibinfo{author}{Grieder, M.} \& \bibinfo{author}{Schmitz, J.}
\newblock \bibinfo{title}{Nudge for {Good}? {Choice} {Defaults} and {Spillover} {Effects}}.
\newblock \emph{\bibinfo{journal}{Frontiers in Psychology}} \textbf{\bibinfo{volume}{10}} (\bibinfo{year}{2019}).
\newblock \urlprefix\url{https://www.frontiersin.org/articles/10.3389/fpsyg.2019.00178}.

\bibitem{van_rookhuijzen_effects_2021}
\bibinfo{author}{Van~Rookhuijzen, M.}, \bibinfo{author}{De~Vet, E.} \& \bibinfo{author}{Adriaanse, M.~A.}
\newblock \bibinfo{title}{The {Effects} of {Nudges}: {One}-{Shot} {Only}? {Exploring} the {Temporal} {Spillover} {Effects} of a {Default} {Nudge}}.
\newblock \emph{\bibinfo{journal}{Frontiers in Psychology}} \textbf{\bibinfo{volume}{12}} (\bibinfo{year}{2021}).
\newblock \urlprefix\url{https://www.frontiersin.org/articles/10.3389/fpsyg.2021.683262}.

\bibitem{lemken_public_2023}
\bibinfo{author}{Lemken, D.}, \bibinfo{author}{Wahnschafft, S.} \& \bibinfo{author}{Eggers, C.}
\newblock \bibinfo{title}{Public acceptance of default nudges to promote healthy and sustainable food choices}.
\newblock \emph{\bibinfo{journal}{BMC Public Health}} \textbf{\bibinfo{volume}{23}}, \bibinfo{pages}{2311} (\bibinfo{year}{2023}).
\newblock \urlprefix\url{https://doi.org/10.1186/s12889-023-17127-z}.

\bibitem{park_choosing_2000}
\bibinfo{author}{Park, C.~W.}, \bibinfo{author}{Jun, S.~Y.} \& \bibinfo{author}{Macinnis, D.~J.}
\newblock \bibinfo{title}{Choosing {What} {I} {Want} versus {Rejecting} {What} {I} {Do} {Not} {Want}: {An} {Application} of {Decision} {Framing} to {Product} {Option} {Choice} {Decisions}}.
\newblock \emph{\bibinfo{journal}{Journal of Marketing Research}} \textbf{\bibinfo{volume}{37}}, \bibinfo{pages}{187--202} (\bibinfo{year}{2000}).
\newblock \urlprefix\url{https://doi.org/10.1509/jmkr.37.2.187.18731}.
\newblock \bibinfo{note}{Publisher: SAGE Publications Inc}.

\bibitem{bosch_tales_2016}
\bibinfo{author}{Bösch, C.}, \bibinfo{author}{Erb, B.}, \bibinfo{author}{Kargl, F.}, \bibinfo{author}{Kopp, H.} \& \bibinfo{author}{Pfattheicher, S.}
\newblock \bibinfo{title}{Tales from the {Dark} {Side}: {Privacy} {Dark} {Strategies} and {Privacy} {Dark} {Patterns}}.
\newblock \emph{\bibinfo{journal}{Proceedings on Privacy Enhancing Technologies}} \textbf{\bibinfo{volume}{2016}}, \bibinfo{pages}{237--254} (\bibinfo{year}{2016}).

\bibitem{sunstein_nudges_2016}
\bibinfo{author}{Sunstein, C.~R.}
\newblock \bibinfo{title}{Nudges {That} {Fail}} (\bibinfo{year}{2016}).
\newblock \urlprefix\url{https://papers.ssrn.com/abstract=2809658}.

\bibitem{de_ridder_nudgeability_2022}
\bibinfo{author}{de~Ridder, D.}, \bibinfo{author}{Kroese, F.} \& \bibinfo{author}{van Gestel, L.}
\newblock \bibinfo{title}{Nudgeability: {Mapping} {Conditions} of {Susceptibility} to {Nudge} {Influence}}.
\newblock \emph{\bibinfo{journal}{Perspectives on Psychological Science}} \textbf{\bibinfo{volume}{17}}, \bibinfo{pages}{346--359} (\bibinfo{year}{2022}).
\newblock \urlprefix\url{https://doi.org/10.1177/1745691621995183}.
\newblock \bibinfo{note}{Publisher: SAGE Publications Inc}.

\bibitem{guido_nudging_2023}
\bibinfo{author}{Guido, A.}, \bibinfo{author}{Tverskoi, D.}, \bibinfo{author}{Gavrilets, S.}, \bibinfo{author}{Sánchez, A.} \& \bibinfo{author}{Andrighetto, G.}
\newblock \bibinfo{title}{Nudging or {Nagging}: {The} {Perils} of {Persuasion}} (\bibinfo{year}{2023}).
\newblock \urlprefix\url{https://papers.ssrn.com/abstract=4404960}.

\bibitem{ghesla_behavioral_2017}
\bibinfo{author}{Ghesla, C.}
\newblock \emph{\bibinfo{title}{Behavioral {Economics} and {Public} {Policy}: {The} {Case} of {Green} {Electricity} {Defaults}}}.
\newblock \bibinfo{type}{Doctoral {Thesis}}, \bibinfo{school}{ETH Zurich} (\bibinfo{year}{2017}).
\newblock \urlprefix\url{https://www.research-collection.ethz.ch/handle/20.500.11850/229934}.
\newblock \bibinfo{note}{Accepted: 2018-02-05T12:18:56Z}.

\bibitem{fehr_field_2011}
\bibinfo{author}{Fehr, E.} \& \bibinfo{author}{Leibbrandt, A.}
\newblock \bibinfo{title}{A field study on cooperativeness and impatience in the {Tragedy} of the {Commons}}.
\newblock \emph{\bibinfo{journal}{Journal of Public Economics}} \textbf{\bibinfo{volume}{95}}, \bibinfo{pages}{1144--1155} (\bibinfo{year}{2011}).
\newblock \urlprefix\url{https://www.sciencedirect.com/science/article/pii/S0047272711000855}.

\bibitem{rustagi_conditional_2010}
\bibinfo{author}{Rustagi, D.}, \bibinfo{author}{Engel, S.} \& \bibinfo{author}{Kosfeld, M.}
\newblock \bibinfo{title}{Conditional {Cooperation} and {Costly} {Monitoring} {Explain} {Success} in {Forest} {Commons} {Management}}.
\newblock \emph{\bibinfo{journal}{Science (New York, N.Y.)}} \textbf{\bibinfo{volume}{330}}, \bibinfo{pages}{961--5} (\bibinfo{year}{2010}).

\bibitem{murphy_social_2013}
\bibinfo{author}{Murphy, R.~O.} \& \bibinfo{author}{Ackermann, K.~A.}
\newblock \bibinfo{title}{Social {Value} {Orientation}: {Theoretical} and {Measurement} {Issues} in the {Study} of {Social} {Preferences}}.
\newblock \emph{\bibinfo{journal}{Personality and Social Psychology Review}}  (\bibinfo{year}{2013}).
\newblock \urlprefix\url{https://journals.sagepub.com/doi/10.1177/1088868313501745}.
\newblock \bibinfo{note}{Tex.ids: murphySocialValueOrientation2013a publisher: SAGE PublicationsSage CA: Los Angeles, CA}.

\bibitem{dave_eliciting_2010}
\bibinfo{author}{Dave, C.}, \bibinfo{author}{Eckel, C.~C.}, \bibinfo{author}{Johnson, C.~A.} \& \bibinfo{author}{Rojas, C.}
\newblock \bibinfo{title}{Eliciting risk preferences: {When} is simple better?}
\newblock \emph{\bibinfo{journal}{Journal of Risk and Uncertainty}} \textbf{\bibinfo{volume}{41}}, \bibinfo{pages}{219--243} (\bibinfo{year}{2010}).
\newblock \urlprefix\url{https://doi.org/10.1007/s11166-010-9103-z}.

\bibitem{eckel_sex_2002}
\bibinfo{author}{Eckel, C.~C.} \& \bibinfo{author}{Grossman, P.~J.}
\newblock \bibinfo{title}{Sex differences and statistical stereotyping in attitudes toward financial risk}.
\newblock \emph{\bibinfo{journal}{Evolution and Human Behavior}} \textbf{\bibinfo{volume}{23}}, \bibinfo{pages}{281--295} (\bibinfo{year}{2002}).
\newblock \urlprefix\url{https://www.sciencedirect.com/science/article/pii/S1090513802000971}.

\bibitem{bogaert_social_2008}
\bibinfo{author}{Bogaert, S.}, \bibinfo{author}{Boone, C.} \& \bibinfo{author}{Declerck, C.}
\newblock \bibinfo{title}{Social value orientation and cooperation in social dilemmas: {A} review and conceptual model}.
\newblock \emph{\bibinfo{journal}{British Journal of Social Psychology}} \textbf{\bibinfo{volume}{47}}, \bibinfo{pages}{453--480} (\bibinfo{year}{2008}).
\newblock \urlprefix\url{https://onlinelibrary.wiley.com/doi/full/10.1348/014466607X244970}.
\newblock \bibinfo{note}{Publisher: John Wiley \& Sons, Ltd}.

\bibitem{lange_social_1998}
\bibinfo{author}{Lange, P. A. M.~V.}, \bibinfo{author}{Vugt, M.~V.}, \bibinfo{author}{Meertens, R.~M.} \& \bibinfo{author}{Ruiter, R. A.~C.}
\newblock \bibinfo{title}{A {Social} {Dilemma} {Analysis} of {Commuting} {Preferences}: {The} {Roles} of {Social} {Value} {Orientation} and {Trust1}}.
\newblock \emph{\bibinfo{journal}{Journal of Applied Social Psychology}} \textbf{\bibinfo{volume}{28}}, \bibinfo{pages}{796--820} (\bibinfo{year}{1998}).
\newblock \urlprefix\url{https://onlinelibrary.wiley.com/doi/abs/10.1111/j.1559-1816.1998.tb01732.x}.
\newblock \bibinfo{note}{\_eprint: https://onlinelibrary.wiley.com/doi/pdf/10.1111/j.1559-1816.1998.tb01732.x}.

\bibitem{fleis_social_2019}
\bibinfo{author}{Fleiß, J.}, \bibinfo{author}{Ackermann, K.~A.}, \bibinfo{author}{Fleiß, E.}, \bibinfo{author}{Murphy, R.~O.} \& \bibinfo{author}{Posch, A.}
\newblock \bibinfo{title}{Social and environmental preferences: measuring how people make tradeoffs among themselves, others, and collective goods}.
\newblock \emph{\bibinfo{journal}{Central European Journal of Operations Research}} \bibinfo{pages}{1--19} (\bibinfo{year}{2019}).

\bibitem{cremer_why_2001}
\bibinfo{author}{Cremer, D.} \& \bibinfo{author}{Lange, P.}
\newblock \bibinfo{title}{Why {Prosocials} {Exhibit} {Greater} {Cooperation} {Than} {Proselfs}: {The} {Roles} of {Social} {Responsibility} and {Reciprocity}}.
\newblock \emph{\bibinfo{journal}{European Journal of Personality}} \textbf{\bibinfo{volume}{15}}, \bibinfo{pages}{S5--S18} (\bibinfo{year}{2001}).

\bibitem{roch_effects_1997}
\bibinfo{author}{Roch, S.~G.} \& \bibinfo{author}{Samuelson, C.~D.}
\newblock \bibinfo{title}{Effects of {Environmental} {Uncertainty} and {Social} {Value} {Orientation} in {Resource} {Dilemmas}}.
\newblock \emph{\bibinfo{journal}{Organizational Behavior and Human Decision Processes}} \textbf{\bibinfo{volume}{70}}, \bibinfo{pages}{221--235} (\bibinfo{year}{1997}).
\newblock \urlprefix\url{https://www.sciencedirect.com/science/article/pii/S0749597897927072}.

\bibitem{montero-porras_fast_2022}
\bibinfo{author}{Montero-Porras, E.}, \bibinfo{author}{Lenaerts, T.}, \bibinfo{author}{Gallotti, R.} \& \bibinfo{author}{Grujic, J.}
\newblock \bibinfo{title}{Fast deliberation is related to unconditional behaviour in iterated {Prisoners}’ {Dilemma} experiments}.
\newblock \emph{\bibinfo{journal}{Scientific Reports}} \textbf{\bibinfo{volume}{12}}, \bibinfo{pages}{20287} (\bibinfo{year}{2022}).
\newblock \urlprefix\url{https://www.nature.com/articles/s41598-022-24849-4}.
\newblock \bibinfo{note}{Number: 1 Publisher: Nature Publishing Group}.

\bibitem{giuliani_joint_2023}
\bibinfo{author}{Giuliani, F.} \emph{et~al.}
\newblock \bibinfo{title}{The joint effect of framing and defaults on choice behavior}.
\newblock \emph{\bibinfo{journal}{Psychological Research}} \textbf{\bibinfo{volume}{87}}, \bibinfo{pages}{1114--1128} (\bibinfo{year}{2023}).
\newblock \urlprefix\url{https://doi.org/10.1007/s00426-022-01726-3}.

\bibitem{buckley_demand_2018}
\bibinfo{author}{Buckley, P.} \& \bibinfo{author}{Llerena, D.}
\newblock \bibinfo{title}{Demand response as a common pool resource game: {Nudges} versus prices} (\bibinfo{year}{2018}).
\newblock \urlprefix\url{https://hal.archives-ouvertes.fr/hal-01704457}.

\bibitem{zlatev_default_2017}
\bibinfo{author}{Zlatev, J.~J.}, \bibinfo{author}{Daniels, D.~P.}, \bibinfo{author}{Kim, H.} \& \bibinfo{author}{Neale, M.~A.}
\newblock \bibinfo{title}{Default neglect in attempts at social influence}.
\newblock \emph{\bibinfo{journal}{Proceedings of the National Academy of Sciences}} \textbf{\bibinfo{volume}{114}}, \bibinfo{pages}{13643--13648} (\bibinfo{year}{2017}).
\newblock \urlprefix\url{https://www.pnas.org/doi/abs/10.1073/pnas.1712757114}.
\newblock \bibinfo{note}{Publisher: Proceedings of the National Academy of Sciences}.

\bibitem{saijo_common-pool_2017}
\bibinfo{author}{Saijo, T.}, \bibinfo{author}{Feng, J.} \& \bibinfo{author}{Kobayashi, Y.}
\newblock \bibinfo{title}{Common-{Pool} {Resources} are intrinsically unstable}.
\newblock \emph{\bibinfo{journal}{International Journal of the Commons}} \textbf{\bibinfo{volume}{11}}, \bibinfo{pages}{597--620} (\bibinfo{year}{2017}).
\newblock \urlprefix\url{http://www.thecommonsjournal.org/articles/10.18352/ijc.692/}.
\newblock \bibinfo{note}{Number: 2}.

\bibitem{joireman_interdependence_1997}
\bibinfo{author}{Joireman, J.~A.}, \bibinfo{author}{Lange, P. A. M.~v.}, \bibinfo{author}{Kuhlman, D.~M.}, \bibinfo{author}{Vugt, M.~v.} \& \bibinfo{author}{Shelley, G.~P.}
\newblock \bibinfo{title}{An {Interdependence} analysis of commuting decisions}.
\newblock \emph{\bibinfo{journal}{European Journal of Social Psychology}} \textbf{\bibinfo{volume}{27}}, \bibinfo{pages}{441--463} (\bibinfo{year}{1997}).
\newblock \urlprefix\url{https://research.vu.nl/en/publications/an-interdependence-analysis-of-commuting-decisions-2}.
\newblock \bibinfo{note}{Publisher: John Wiley and Sons Ltd}.

\bibitem{kobis_bad_2021}
\bibinfo{author}{Köbis, N.}, \bibinfo{author}{Bonnefon, J.-F.} \& \bibinfo{author}{Rahwan, I.}
\newblock \bibinfo{title}{Bad machines corrupt good morals}.
\newblock \emph{\bibinfo{journal}{Nature Human Behaviour}} \textbf{\bibinfo{volume}{5}}, \bibinfo{pages}{679--685} (\bibinfo{year}{2021}).
\newblock \urlprefix\url{https://www.nature.com/articles/s41562-021-01128-2}.
\newblock \bibinfo{note}{Bandiera\_abtest: a Cg\_type: Nature Research Journals Number: 6 Primary\_atype: Reviews Publisher: Nature Publishing Group Subject\_term: Computer science;Human behaviour;Science, technology and society Subject\_term\_id: computer-science;human-behaviour;science-technology-and-society}.

\bibitem{european_parlament_regulation_2022}
\bibinfo{author}{{European Parlament}}.
\newblock \bibinfo{title}{Regulation ({EU}) 2022/2065 of the {European} {Parliament} and of the {Council} of 19 {October} 2022 on a {Single} {Market} {For} {Digital} {Services} and amending {Directive} 2000/31/{EC} ({Digital} {Services} {Act}) ({Text} with {EEA} relevance)} (\bibinfo{year}{2022}).
\newblock \urlprefix\url{http://data.europa.eu/eli/reg/2022/2065/oj/eng}.
\newblock \bibinfo{note}{Legislative Body: EP, CONSIL}.

\bibitem{european_parlament_regulation_2016}
\bibinfo{author}{{European Parlament}}.
\newblock \bibinfo{title}{Regulation ({EU}) 2016/679 of the {European} {Parliament} and of the {Council} of 27 {April} 2016 on the protection of natural persons with regard to the processing of personal data and on the free movement of such data, and repealing {Directive} 95/46/{EC} ({General} {Data} {Protection} {Regulation}) ({Text} with {EEA} relevance)} (\bibinfo{year}{2016}).
\newblock \urlprefix\url{http://data.europa.eu/eli/reg/2016/679/oj/eng}.
\newblock \bibinfo{note}{Legislative Body: EP, CONSIL}.

\bibitem{european_parlament_directive_2009}
\bibinfo{author}{{European Parlament}}.
\newblock \bibinfo{title}{Directive 2009/125/{EC} of the {European} {Parliament} and of the {Council} of 21 {October} 2009 establishing a framework for the setting of ecodesign requirements for energy-related products (recast) ({Text} with {EEA} relevance)} (\bibinfo{year}{2009}).
\newblock \urlprefix\url{http://data.europa.eu/eli/dir/2009/125/oj/eng}.

\bibitem{murphy_measuring_2011}
\bibinfo{author}{Murphy, R.~O.}, \bibinfo{author}{Ackermann, K.~A.} \& \bibinfo{author}{Handgraaf, M.}
\newblock \bibinfo{title}{Measuring {Social} {Value} {Orientation}}.
\newblock \bibinfo{type}{{SSRN} {Scholarly} {Paper}} \bibinfo{number}{ID 1804189}, \bibinfo{institution}{Social Science Research Network}, \bibinfo{address}{Rochester, NY} (\bibinfo{year}{2011}).
\newblock \urlprefix\url{https://papers.ssrn.com/abstract=1804189}.

\bibitem{bravo_agents_2011}
\bibinfo{author}{Bravo, G.}
\newblock \bibinfo{title}{Agents’ beliefs and the evolution of institutions for common-pool resource management}.
\newblock \emph{\bibinfo{journal}{Rationality and Society}} \textbf{\bibinfo{volume}{23}}, \bibinfo{pages}{117--152} (\bibinfo{year}{2011}).
\newblock \urlprefix\url{https://doi.org/10.1177/1043463110387268}.
\newblock \bibinfo{note}{Publisher: SAGE Publications Ltd}.

\bibitem{lange_pursuit_1999}
\bibinfo{author}{Lange, P.~V.}
\newblock \bibinfo{title}{The {Pursuit} of {Joint} {Outcomes} and {Equality} in {Outcomes}: {An} {Integrative} {Model} of {Social} {Value} {Orientation}}.
\newblock \emph{\bibinfo{journal}{Journal of Personality and Social Psychology}} \textbf{\bibinfo{volume}{77}}, \bibinfo{pages}{337--349} (\bibinfo{year}{1999}).

\bibitem{liebrand_value_1986}
\bibinfo{author}{Liebrand, W.~B.}, \bibinfo{author}{Wilke, H.~A.}, \bibinfo{author}{Vogel, R.} \& \bibinfo{author}{Wolters, F.~J.}
\newblock \bibinfo{title}{Value {Orientation} and {Conformity}: {A} {Study} {Using} {Three} {Types} of {Social} {Dilemma} {Games}}.
\newblock \emph{\bibinfo{journal}{Journal of Conflict Resolution}} \textbf{\bibinfo{volume}{30}}, \bibinfo{pages}{77--97} (\bibinfo{year}{1986}).
\newblock \urlprefix\url{https://doi.org/10.1177/0022002786030001006}.
\newblock \bibinfo{note}{Publisher: SAGE Publications Inc}.

\bibitem{wood_fast_2011}
\bibinfo{author}{Wood, S.~N.}
\newblock \bibinfo{title}{Fast stable restricted maximum likelihood and marginal likelihood estimation of semiparametric generalized linear models}.
\newblock \emph{\bibinfo{journal}{Journal of the Royal Statistical Society: Series B (Statistical Methodology)}} \textbf{\bibinfo{volume}{73}}, \bibinfo{pages}{3--36} (\bibinfo{year}{2011}).
\newblock \urlprefix\url{https://onlinelibrary.wiley.com/doi/abs/10.1111/j.1467-9868.2010.00749.x}.
\newblock \bibinfo{note}{\_eprint: https://onlinelibrary.wiley.com/doi/pdf/10.1111/j.1467-9868.2010.00749.x}.

\bibitem{wood_mgcv_2023}
\bibinfo{author}{Wood, S.}
\newblock \bibinfo{title}{mgcv: {Mixed} {GAM} {Computation} {Vehicle} with {Automatic} {Smoothness} {Estimation}} (\bibinfo{year}{2023}).
\newblock \urlprefix\url{https://cran.r-project.org/web/packages/mgcv/index.html}.

\end{thebibliography}

\section{Acknowledgements}

E.M. and T.L benefit from the support by the Flemish Government through the AI Research Program and by TAILOR, a project funded by the EU Horizon 2020 research and innovation program under GA No 952215.
T.L. is furthermore supported by the F.N.R.S. projects with grant number 31257234 and 40007793,  the F.W.O. project with grant no. G.0391.13N, and the Service Public de Wallonie Recherche under grant n\textdegree\ 2010235–ARIAC by DigitalWallonia4.ai.
E.F.D. is supported by an F.N.R.S Chargé de Recherche position, grant number 40005955.

\section{Author information}

\subsection{Contributions}
E.M.P., R.S. and E.F.D. designed the study and wrote the paper; E.M.P. implemented the experimental platform, collected data, and performed the analysis. T.L. wrote and reviewed the paper. All authors edited and approved the final version.

\subsection{Corresponding authors}
Eladio Montero-Porras
Elias Fernández Domingos

\section*{Ethics declarations}
\subsection{Competing interests}
The authors declare no competing interests.

\section{Source data}
The source dataset, the code used for the analysis and to reproduce the figures can be found in Zenodo 
\newpage

\begin{appendices}

\section{Experimental design}

\subsection{Deviations from pre-registration}

The pre-registered experiment, shown in this paper (\href{https://osf.io/jg2sa}{https://osf.io/jg2sa}), was performed from May 2023 until August 2023. 

We made the last changes to the pre-registration in March 2023. In these changes, we clarified the experimental hypotheses and uploaded the latest documents before starting the experiments. Here are some updates we did after that:

\begin{itemize}
    \item \textbf{Sample size rationale}: Our initial power analysis was based on the work by Pichert and Katsikopoulos \cite{pichert_green_2008} on default adoption of green energy options, this was done in July 2022. However, in later iterations, we adapted our calculations and effect size based on a meta-analysis done by Jachimowicz et al. \cite{jachimowicz_when_2019}. This meta-analysis included the effect size of 58 studies, which we found was more robust. However, by mistake, we did not upload the updated file to OSF, where we estimated a required sample size of about 69 participants per treatment, instead of the 54 needed according to the initial analysis. This new power analysis was performed on January 2023, before the experiments took place. The correct file was updated and uploaded in OSF in February 2024. Moreover, we adapted the effect size to $F=0.25$, $\alpha = 0.05$ and $1 - \beta = 0.95$. Additionally, we assumed a low correlation among repeated measures ($0.1$). In every treatment, we chose a significantly larger sample size than the minimum required, to account for estimation errors and ensure that we have sufficient power. 
    \item \textbf{Repeated Measures ANOVA}: We switched from repeated measures ANOVA (RMANOVA) to Generalised Additive Mixed Regression models due to the non-normal distribution in time periods, violating RMANOVA assumptions.
    \item \textbf{Treatment names:} Changed from 'High or Fair Extraction' (HE/FE) to 'Self-serving or Prosocial', for clarity.
\end{itemize}

\section{Experiment metadata}\label{secA1}

Table \ref{tab:status} shows the distribution of the participants according to their final status in the experiment. \emph{Completed} refers to the participants who reached the end of the experiment, while \emph{incomplete} refers to the people who left before the end of the experiment, without cashing any payoff (normally people that withdrew consent or just did not finish the first two tasks). \emph{Failed test} were participants who completed the first two tasks but did not pass the comprehension test for task 3. Status of \emph{ungrouped} means that they completed the first two tasks and the comprehension test, but were waiting for 15 minutes or more for a group of 4 to be formed. \emph{Lost focus} is all participants who were waiting to be grouped but either left or lost the focus of their browser tab when they were being grouped. \emph{Timeout} refers to people who were excluded because they did not make a decision within 2 minutes in a given round of task 3.
All participants except for the ones with status \emph{incomplete} were paid for their time (6 pounds per hour) and a bonus for each task they managed to complete.

Of the participants who completed the experiment, some groups had to be removed, specifically those who had a group member who was timed-out. Thus, our final sample size used in the results shown in this manuscript is 412 participants across all treatments. In the Self-serving treatment: 156, Pro-social: 156 and Control: 100. 50\% of them were Female. The mean age of participants was 28.77 years old (SD = 9.02 years). The average time taken for the experiment was 28 min (SD = 9.10 minutes). Main nationalities are South African = 21\%, Portuguese = 16\% and Polish = 15\%. Moreover, 30\% of them were employed full-time, 13\% part-time employees and 15\% unemployed. All participants were fluent in English, had at least 10 successful submissions on Prolific and had a 99-100\% success rate in their submissions. None of them could participate more than once in the experiment as past sessions’ participants were filtered out. 

\begin{table}[!ht]
\begin{tabular}{ccc}
\textbf{Status} & \textbf{No. participants} & \textbf{\%}    \\
Completed       & 494                       & 70\%         \\
Failed test   & 74                        & 10.4\%         \\
Incomplete      & 71                        & 10.0\%         \\
Ungrouped       & 49                        & 7\%          \\
Timeout         & 19                        & 2.6\%          \\
Lost focus      & 2                         & 0.2\%          \\
                & \textbf{709}              & \textbf{100\%}
\end{tabular}
\caption{Number of participants per status in the experiment.} 
\label{tab:status}
\end{table}

\newpage

\begin{table}[!h]
\centering

\begin{tabular}{l l l l l} 
 & \textbf{Estimate} & \textbf{Std. Error} & \textbf{z value} & \textbf{Pr($>$|z|)} \\
(Intercept) & -43.9800 & 60860 & -0.0010 & 0.9990 \\
svo\_score & -0.0126 & 0.0320 & -0.3930 & 0.6940 \\
gamble\_choice & 0.3874 & 0.2460 & 1.5750 & 0.1150 \\
Nationality$^{***}$ & -3.8309 & 60957 & 0.0000 & 1.0000 \\
Age\_x & -0.0010 & 0.0466 & -0.0220 & 0.9820 \\
Sex\_xFemale & 17.1400 & 31200 & 0.0010 & 1.0000 \\
Sex\_xMale & -1.3440 & 31320 & 0.0000 & 1.0000 \\
Sex\_xPrefer not to say & 15.0400 & 57900 & 0.0000 & 1.0000 \\
Ethnicity\_simplified\_xBlack & 20.7300 & 15670 & 0.0010 & 0.9990 \\
Ethnicity\_simplified\_xDATA\_EXPIRED & 1.8900 & 50680 & 0.0000 & 1.0000 \\
Ethnicity\_simplified\_xMixed & 3.6620 & 17890 & 0.0000 & 1.0000 \\
Ethnicity\_simplified\_xOther & 3.0780 & 20190 & 0.0000 & 1.0000 \\
Ethnicity\_simplified\_xWhite & 19.8500 & 15670 & 0.0010 & 0.9990 \\
Total\_approvals\_x & 0.0018 & 0.0017 & 1.0850 & 0.2780 \\

\end{tabular}
\caption{Results of the logistic regression made to test if the 21 participants who dropped out did so because of their results in previous tasks (variables \textit{svo\_score} and \textit{gamble\_choice}) or their demographic information. \textbf{$***$:} due to the number of options, the \textit{Nationality} variable was grouped as a single variable and the estimates were averaged.}
\label{tab:dropouts}
\end{table}

\begin{figure}[!ht]%
\centering
\includegraphics[width=\linewidth]{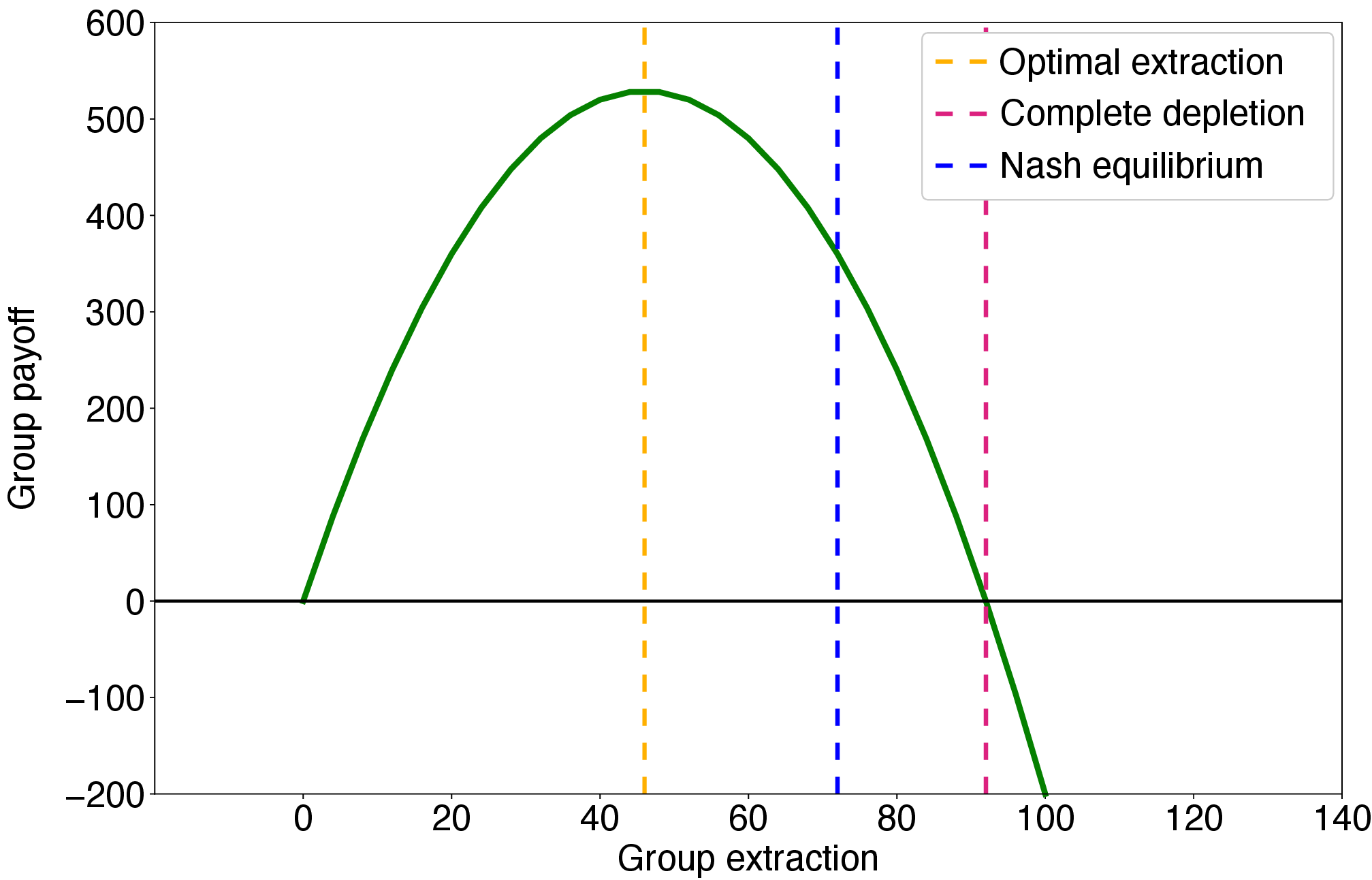}
\caption{Group payoff as a result of a given group extraction. As shown in the green line, the maximum return the CPRD can provide is when participants extract 46 tokens as a group (average 11.5 tokens per player), which is shown with the yellow dashed line. If participants extract more or less than that quantity as a group, they will be under or over extracting the resource, respectively. The pink line shows that if participants extract 92 tokens as a group (average 23 tokens per player) they will all receive zero ECUs. }\label{fig:group_extraction_payoff}
\end{figure}

\subsection{Default picking and self-assessment}\label{sec:default_assessment}

At the end of the experiment, we asked participants to self-assess the influence the default value had on their extractions, this question was open-ended and was only shown to participants in the Self-serving and Pro-social treatments. The question was: \textit{``Did the presence of a default value in the first 5 rounds affected your desired extraction?''.} 

As this question was open-ended, we manually formatted each participant's response and classified ``Not affected'' those who responded either ``\textit{No}'', ``\textit{Not really}'' or something similar the denoted a clear negative response to the question. We did not label as ``Not affected'' those who responded a more vague reaction, such as: ``\textit{I don't think so, only in the first one}'' or ``\textit{Only slightly}''. We were this strict in the negative responses because prior work has showed that individuals often underestimate the influence of default choices on their decisions \cite{zlatev_default_2017}, and we wanted to test this new hypothesis. In total, we classified $161$ participants as ``Not affected'' and $151$ as ``Affected''. 

We analysed the extractions of the self-assessed ``Not affected'' with the extractions of both the ``Affected'' and the participants in the Control treatment, where no default value was displayed.  

Comparing the mean extraction in the Control treatment $\overline{x} = 15.9280$, 95\% CI = $[15.3137, 16.5423]$, with those who self-assessed as ``Not affected'' is significantly different in both cases, in the Pro-social treatment (KS test $=0.097398,$ p-value $= 0.02707$) and in the Self-serving treatment (KS test $= 0.19523$, p-value $ < 0.0001$). 

The mean extraction of the ``Not affected'' participants in the Pro-social treatment was $\overline{x} = 15.3855$, 95\% CI = $[14.7947, 15.9763]$ while the ``Affected'' was $\overline{x} = 14.8164$, 95\% CI = $[14.2624, 15.3704]$. This difference is not significant (KS test: $= 0.080607$, p-value $= 0.1603$). In the Self-serving treatment, ``Not affected'' extracted $\overline{x} = 18.1462$, 95\% CI = $[17.4712, 18.8211]$ and the ``Affected'' $\overline{x} = 19.4051$, 95\% CI = $[18.7876, 20.0227]$ and this difference is significant (KS test: $= 0.14359, p-value = 0.0006$).

\clearpage

\subsection{Tasks 1 and 2 supplementary material}

\begin{figure}[!ht]%
\centering
\includegraphics[width=\linewidth]{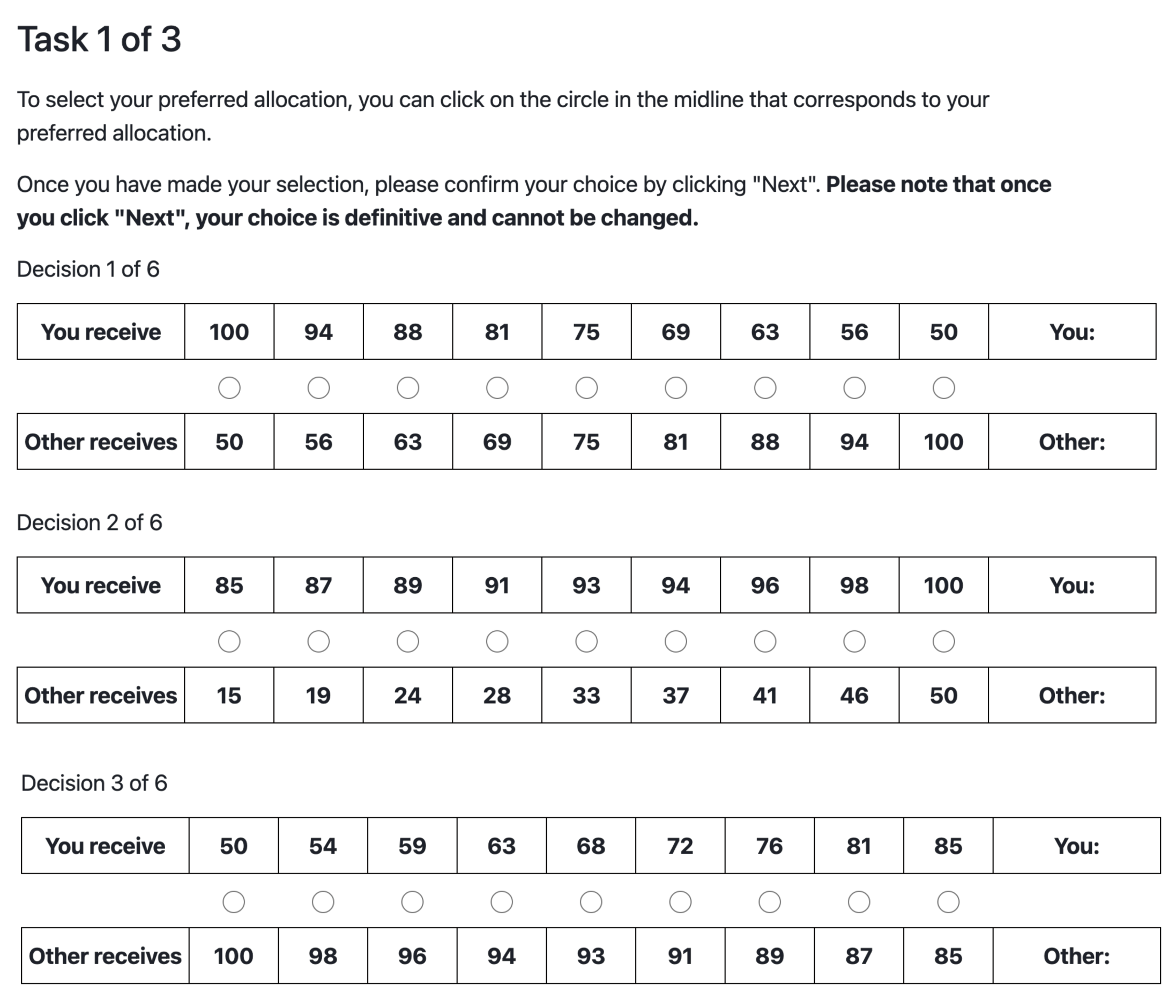}
\caption{Sample of three out of six decisions participants had to make in the first task. Each decision was shown individually, to not crowd the user's interface. The decisions were taken from Murphy's et al. Slider Measure, with the six main measures \cite{murphy_measuring_2011}.}
\label{fig:svo_sample_decisions}
\end{figure}

\begin{figure}[!ht]%
\centering
\includegraphics[width=\linewidth]{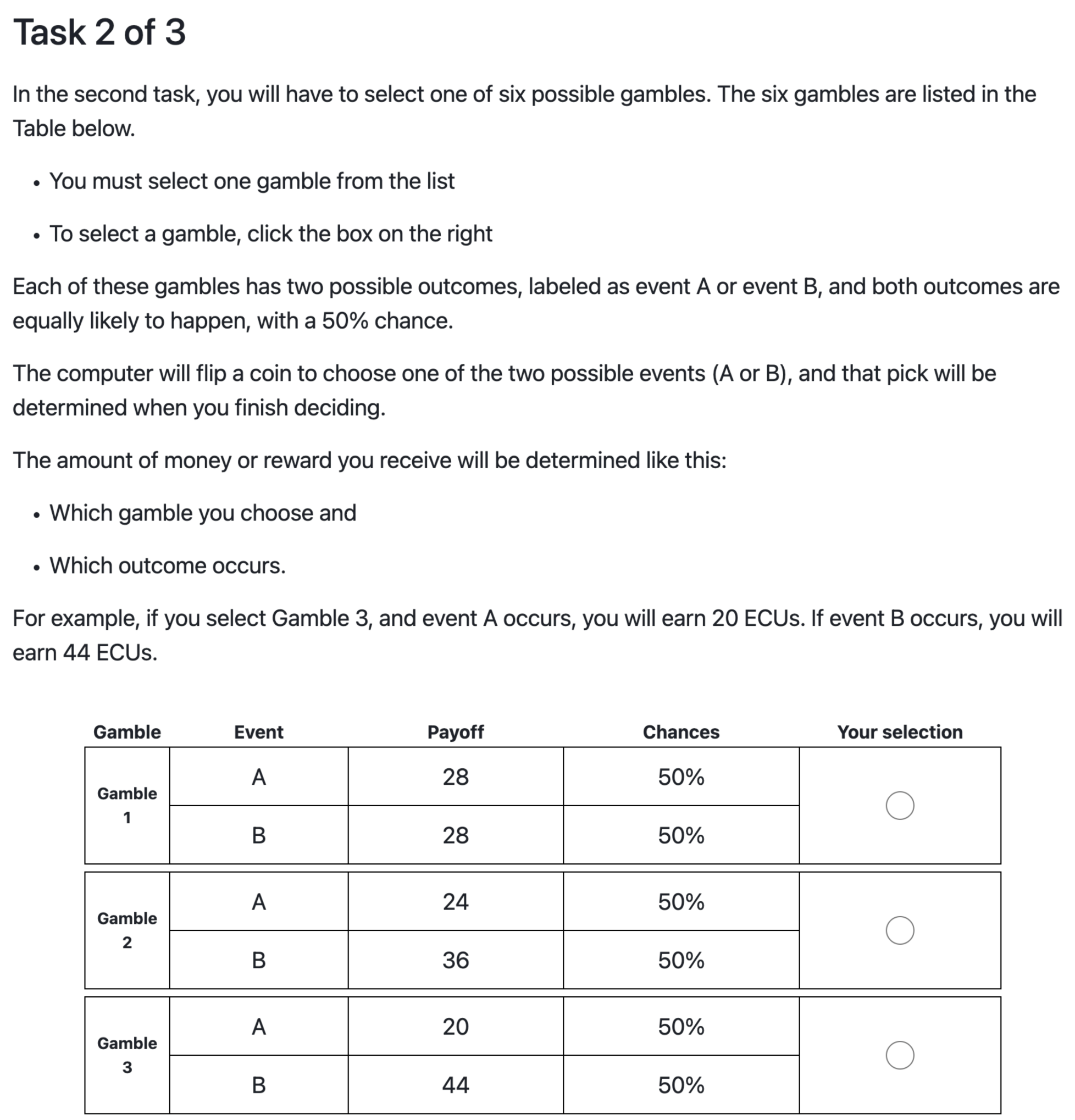}
\caption{Sample of three out of six options participants had for the second task. In this case, all options were shown in the same table for comparison. The options were taken from the measure developed by Dave et al \cite{dave_eliciting_2010}. }
\label{fig:risk_sample_decisions}
\end{figure}

\begin{figure}[!ht]%
\centering
\includegraphics[width=\linewidth]{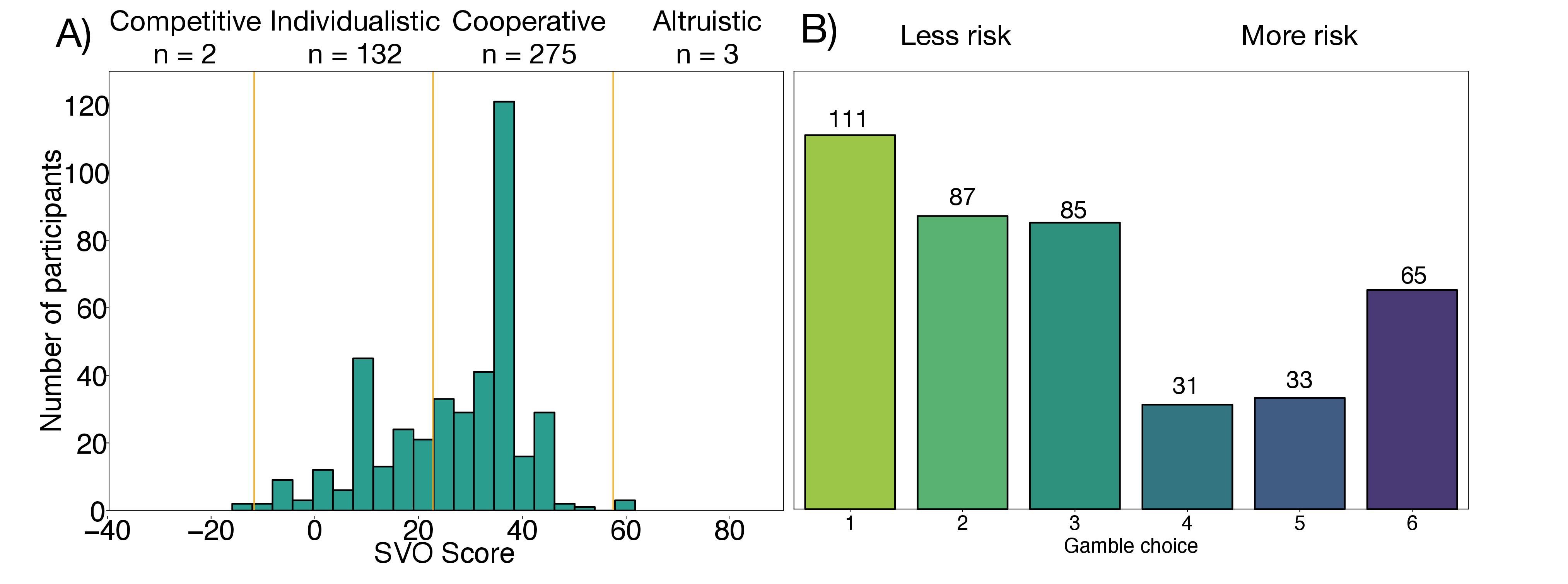}
\caption{Left: Distribution of participants according to their SVO Score (task 1). This test measures a participant’s resource allocation preferences. Values near 0 represent more individualistic preferences, while values towards 90 represent more altruistic preferences. Right: Distribution of participants in the Task 2 (Risk assessment). In this task, participants had to choose between 6 gambles, where 1 was the less risky and 6 was the riskiest one.}
\label{fig:svo_gamble}
\end{figure}

\begin{figure}[h]%
\centering
\includegraphics[width=\linewidth]{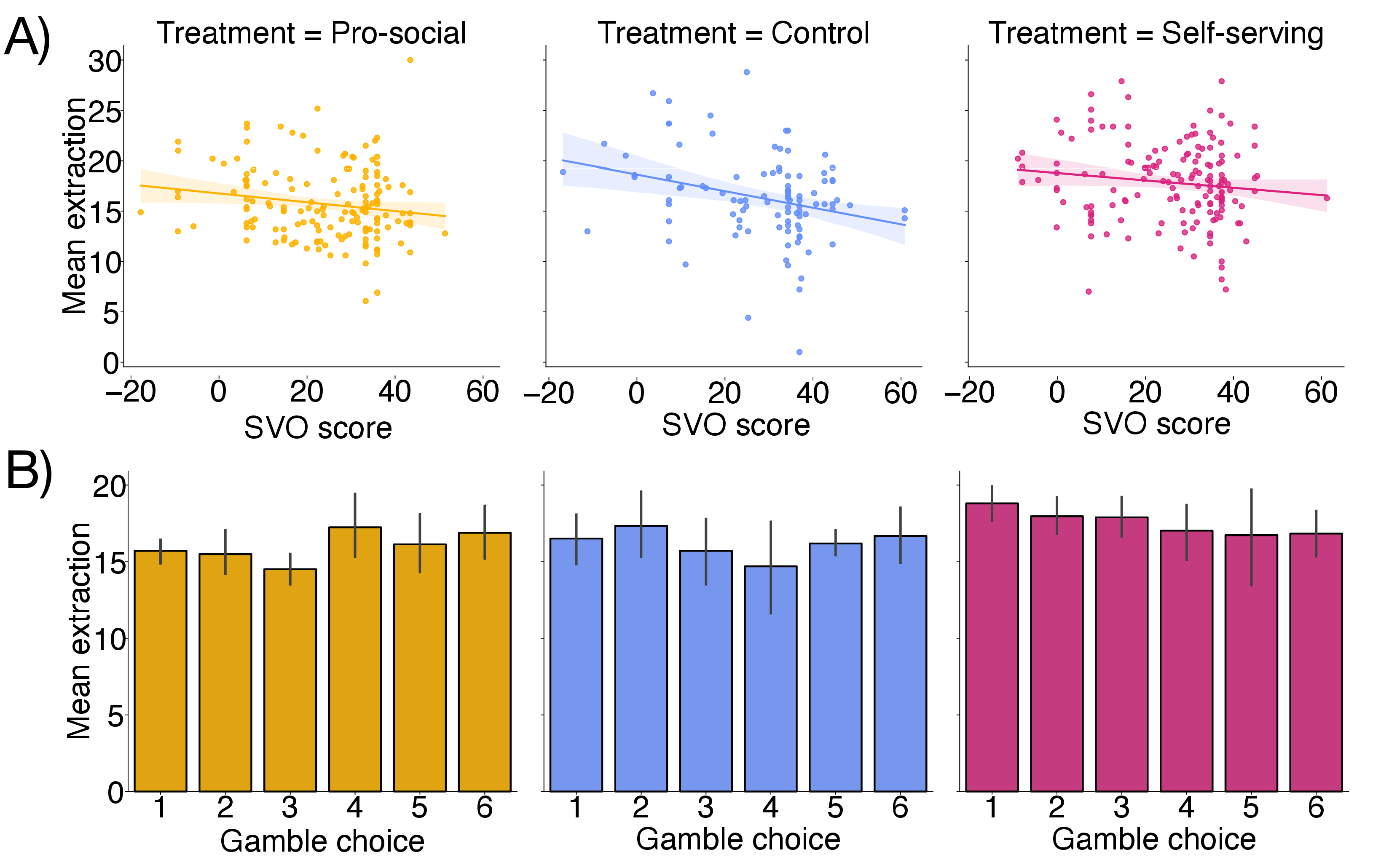}
\caption{\textbf{A}: relationship between participant’s SVO score in task 1 and their mean extraction in all rounds. The line across represents the linear fit and the shades the 95\% confidence interval. \textbf{B:} Mean extraction of the participants according to their gamble choices in Task 2 of the experiment.}\label{fig:svo_gamble_extraction}
\end{figure}

\begin{figure}[h]%
\centering
\includegraphics[width=\linewidth]{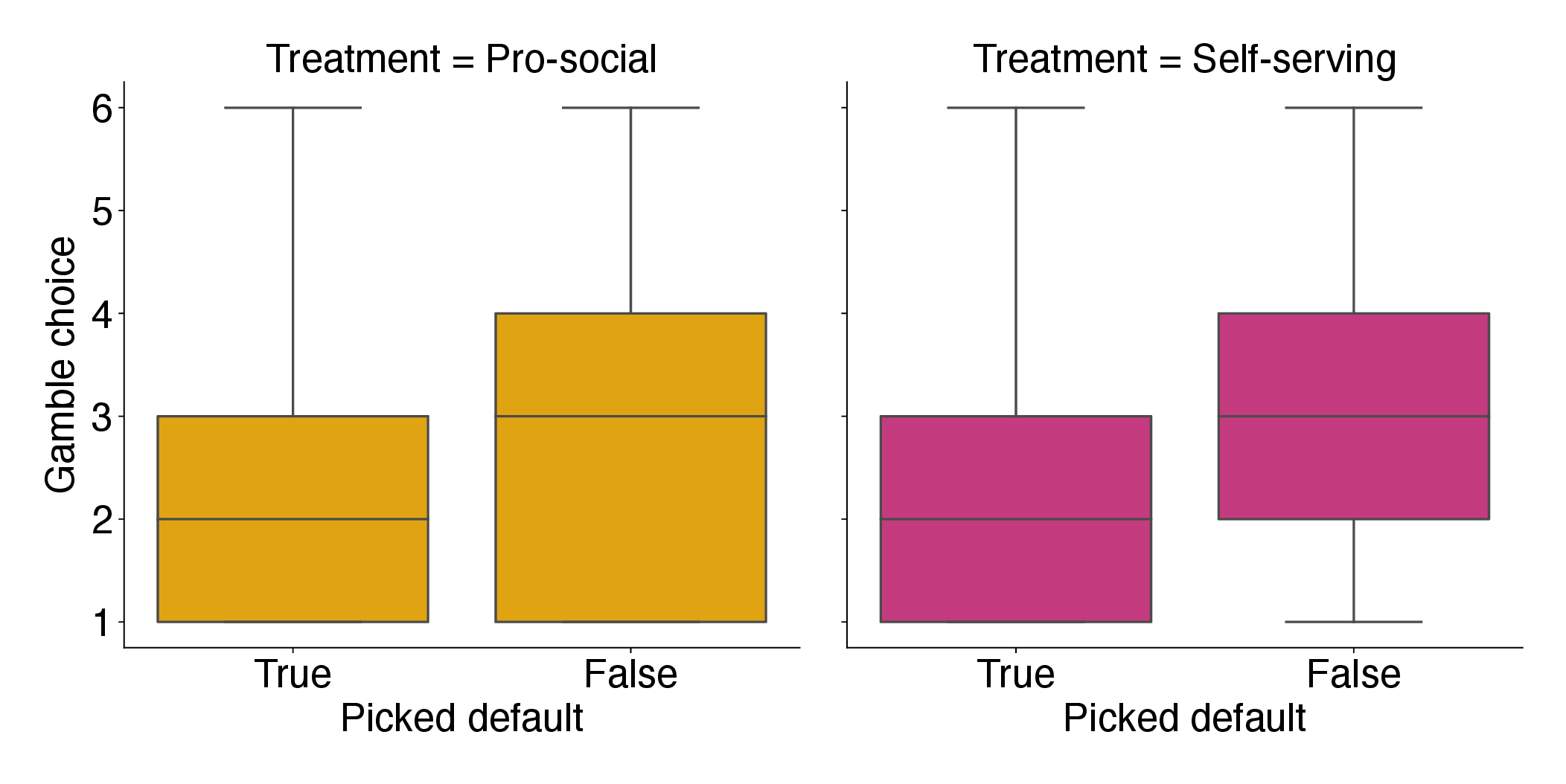}
\caption{\textbf{Gamble choice and extracting the default option}. We divided participants who extracted exactly the default value provided (marked as True) and those who did not (marked as False), in any round, so they could have picked the default in one round and change their extraction in the next one. In the Pro-social treatment, 19\% ($147$ out of $780$ decisions) of the extractions were the default option in the first five rounds, compared to 24\% ($184$ out of $780$ decisions) in the Self-serving treatment.}\label{fig:gamble_default}
\end{figure}

Figure \ref{fig:gamble_default} shows the gamble choice of those who picked the approximate default in task 3 and those who picked another value. As shown in the figure, the gamble choice of default extractors is lower in both Pro-social (KS: 0.1224, p-value: 0.0177)
and Self-serving treatment (KS: 0.1266, p-value: 0.0048). This means that risk-averse individuals picked the default more often than risk-seekers.

\clearpage

\subsection{Results from the Mixed effects model}

\begin{table}[!h]
\centering

\begin{tabular}{cccccc}
Parametric
  coefficients &  &  &  &  &  \\
 & \textbf{Estimate} & \textbf{Std. Error} & \textbf{t value} & \textbf{Pr(\textless|t|)} &  \\
\textbf{(Intercept)} & 7.7650 & 0.3655 & 21.2470 & \textless 2e-16 & *** \\
\textbf{Treatment:Pro-social} & -0.7952 & 0.5307 & -1.4980 & 0.1341 &  \\
\textbf{Treatment:Self-serving} & 1.1464 & 0.5318 & 2.1560 & 0.0312 & * \\
\textbf{extraction\_last\_round} & 0.0823 & 0.0148 & 5.5640 & 0.0001 & *** \\
\textbf{others\_extraction\_last\_round} & 0.0064 & 0.0086 & 0.7470 & 0.4549 &  \\
\textbf{svo\_score} & 0.2927 & 0.0136 & 21.4500 & \textless 2e-16 & *** \\
\textbf{gamble\_choice} & -0.1215 & 0.1162 & -1.0460 & 0.2958 &  \\

\end{tabular}
\end{table}

\begin{table}[!h]
\centering

\begin{tabular}{cccccc}

 Approximate significance of smooth terms&  &  &  &  &  \\
 & \textbf{edf} & \textbf{Ref.df} & F & \textbf{p-value} &  \\
\textbf{round*Treatment:Control} & 1 & 1 & 1.971 & 0.16045 &  \\
\textbf{round*Treatment:Pro-social} & 2.1844 & 2.7316 & 5.1910 & 0.00238 & ** \\
\textbf{round*Treatment:Self-serving} & 6.7837 & 7.9327 & 9.8370 & \textless 2e-16 & *** \\
\textbf{svo\_score*Treatment:Control} & 0.8471 & 0.8471 & 193.7410 & \textless 2e-16 & *** \\
\textbf{svo\_score*Treatment:Pro-social} & 0.8470 & 0.8471 & 216.9680 & \textless 2e-16 & *** \\
\textbf{svo\_score*Treatment:Self-serving} & 0.8470 & 0.8470 & 206.5900 & \textless 2e-16 & *** \\
\textbf{player\_id} & 333.0459 & 405.0000 & 2.6550 & < & *** \\

\end{tabular}
\caption{Results from the Mixed-effects linear model of the variable extraction. The upper table indicates the parametric coefficients which are the fixed effects of the model. The table below shows the smoothing terms, which assumes a non-linear interaction between two variables, and the last row indicates the random effects we included by player, to account for individual heterogeneity and other effects that might not be captured by our model.}
\label{tab:mixed_ef}
\end{table}

\newpage

\begin{figure}[h]%
\centering
\includegraphics[width=\linewidth]{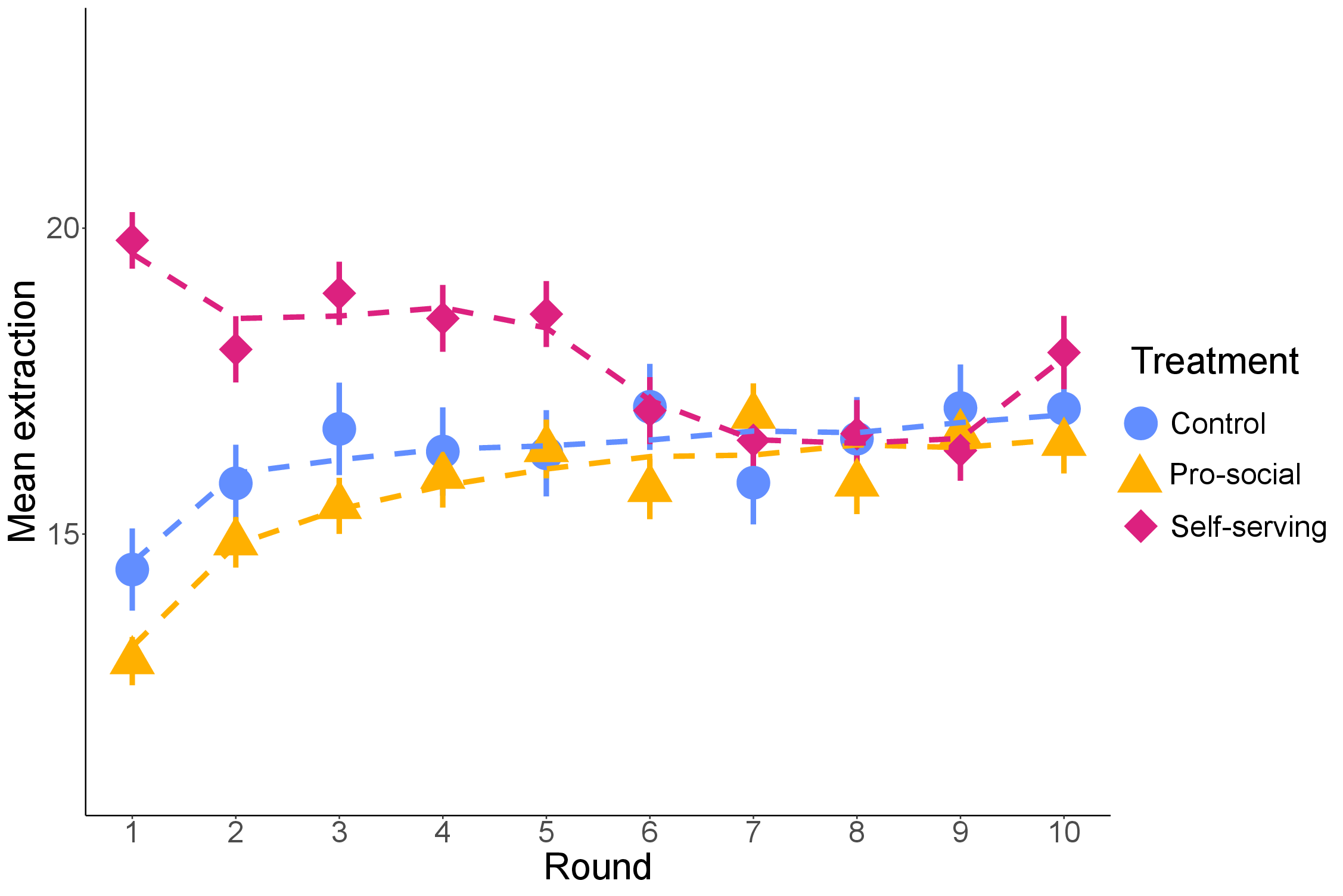}
\caption{Mean extraction by treatment, with the fit given by the Mixed-effects regression model. For more information about this model, see the Methods in section \ref{sec:methods_mixed_models}. We accounted for multiple intercepts and effect of the round depending on the Treatment, SVO and risk assessment, plus a random effect for individual heterogeneity. }
\label{fig:line_treatment_fit}
\end{figure}

\end{appendices}
\newpage

\end{document}